\documentclass
[byrevtex,superscriptaddress,rmp,floatfix,twocolumn,10pt]{revtex4}%
\usepackage{amsfonts}
\usepackage{amsmath}
\usepackage{amssymb}
\usepackage{graphicx}
\usepackage[colorlinks,linktocpage,linkcolor=black,citecolor=black,]%
{hyperref}
\usepackage{version}%
\providecommand{\U}[1]{\protect\rule{.1in}{.1in}}
\begin{document}
\preprint{ }
\title{\textit{Colloquium: }Coherent Diffusion of Polaritons in Atomic Media}
\author{O. Firstenberg}
\affiliation{Department of Physics, Technion-Israel Institute of Technology, Haifa 32000, Israel}
\affiliation{Department of Physics, Harvard University, Cambridge, MA 02138, USA}
\author{M. Shuker}
\affiliation{Department of Physics, Technion-Israel Institute of Technology, Haifa 32000, Israel}
\author{A. Ron}
\affiliation{Department of Physics, Technion-Israel Institute of Technology, Haifa 32000, Israel}
\author{N. Davidson}
\affiliation{Department of Physics of Complex Systems, Weizmann Institute of Science,
Rehovot 76100, Israel}

\begin{abstract}
Coherent diffusion pertains to the motion of atomic dipoles experiencing
frequent collisions in vapor while maintaining their coherence. Recent
theoretical and experimental studies on the effect of coherent diffusion on
key Raman processes, namely Raman spectroscopy, slow polariton propagation,
and stored light, are reviewed in this Colloquium.

\end{abstract}
\maketitle
\tableofcontents

\section{Introduction}

Coherent Raman processes, in which two or more electromagnetic modes
resonantly dress and excite an atomic-like system, provide a powerful
interface between light and matter. They are potentially a cornerstone for
future quantum information schemes and quantum-technology sensors, allowing
the initialization, control, and monitoring of the quantum state of either the
material or the light. Various Raman processes have been studied to date,
namely, coherent population trapping (CPT) \cite{Arimondo96}, nonlinear
magneto-optical rotation (NMOR) \cite{BudkerRMP2002},
electromagnetically-induced transparency (EIT) \cite{FleischhauerRMP2005}, and
slow and stored light \cite{LukinRMP2003,PolzikRMP2010}. These were all first
demonstrated in a hot atomic vapor, perhaps the epitome of quantum-optics
systems, combining high optical depth, low relaxation rates, and weak
atom-atom interactions with the simplicity of both the experiments and the
theoretical modeling. Indeed --- from the pioneering work of
\textcite{AlzettaNuCimen1976}
and
\textcite{ArimondoNuCimenLett1976}
on dark resonances, through later manifestations of elaborate Raman processes
and dark-state polaritons \cite{HarrisPT1997, BudkerPRL1999, PhillipsPRL2001},
and to state-of-the-art magnetometers, gyrometers, and miniature atomic clocks
\cite{BudkerRomalisNP2007,RomalisPRL2011,knappe:1460}\ --- thermal atomic
media have been at the frontier of experimental progress.

Two profound mechanisms underlie the dynamics of coherent processes in vapor:
the continuous thermal motion of the atoms and the collisions amongst
themselves and with the walls of the vapor cell. Collisions damage the
internal atomic quantum state and set an upper limit on the coherence time of
the system. Although a record coherence time of one minute was recently
obtained by
\textcite{BudkerPRL2010}
with an anti-relaxation coating of the inner glass walls, it is often
desirable to add a foreign buffer-gas into the cell to delay the active atoms
from leaving the illuminated region and approaching the walls
\cite{HapperRMP1972}.\textsl{ }Selected species, such as noble gases or
nitrogen molecules, have been known for many years to preserve the
ground-state coherence of alkali-metal atoms upon collisions
\cite{WalkerPRA1989}. Buffered cells are now commonly used in coherent Raman
experiments \cite{HemmerIEEE1995,GrafPRA1995,MeschedePRA1997}.

Frequent velocity-changing collisions, although preserving the coherence,
affect the atomic motion and modify the light-matter interaction. The original
descriptions, by C. Doppler, W. Voigt, and others, of the interplay between a
moving radiator and the electromagnetic field were augmented by R. H.
\textcite{Dicke1953}
to incorporate frequent changes in the radiator velocity. Dicke predicted
that, when collisions dominate, the Doppler-broadened spectrum of a thermal
gas will be narrowed. The Dicke effect is closely related to motional
narrowing in NMR, treated previously in the pioneering paper by
\textcite{BloembergenPR1948}%
. Subsequently,
\textcite{GalatryPR1961}
formulated the spectral lineshape of a thermal atom undergoing frequent
collisions in a buffer gas. Nevertheless it was only in 2003 when a signature
of Dicke narrowing was detected in the optical regime \cite{Dutier2003},
because of the fundamental requirement that the mean free-path between
collisions $\Lambda$ be much smaller the wavelength $\lambda=2\pi
/|\mathbf{q}|$, where $\mathbf{q}$ is the wavevector.

\begin{figure}
[ptb]
\begin{center}
\includegraphics[
height=4.3184cm,
width=6.699cm
]%
{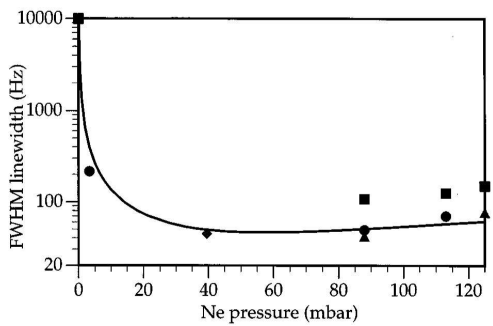}%
\caption{Measured and calculated linewidths of a hyperfine dark resonance in
rubidium, demonstrating the reduction in both the Doppler and transit-time
broadenings with increasing pressure of neon buffer-gas. Laser intensities are
($\square$) 17, ($\triangle$) 11, ($\bigcirc$) 6, and ($\Diamond$) 1
mW/cm$^{2}.$ From \textcite{MeschedePRA1997}.}%
\label{fig_mechede}%
\end{center}
\end{figure}

In Raman processes, however, the relevant wavevector for the Doppler and the
Dicke mechanisms is due to the difference between the two fields involved
$\mathbf{k}=\mathbf{q}-\mathbf{q}_{c}$, leading to the \emph{residual} Doppler
and Dicke effects \cite{Cyr1993}. Broadening is avoided only in the so-called
Doppler-free arrangement, in which one light beam excites an atom and a
collinear beam of the same frequency de-excites it, yielding $\mathbf{k}%
=\mathbf{q}-\mathbf{q}=0$. However, in general either a small angular
deviation or a small frequency difference between the two beams yield a
non-zero Raman wavelength $\lambda_{R}=2\pi/|\mathbf{k}|$ as small as a
micrometer or as large as a centimeter, which affects the process. Residual
Dicke narrowing of a Raman transition at the GHz frequency range is therefore
readily obtained at moderate buffer-gas pressures, as exemplified in Fig.
\ref{fig_mechede} for a Raman dark-resonance. Correspondingly, general
multimode light fields that span a spectrum in $\mathbf{k-}$space exhibit a
generalized motional effect.%

From the spatial viewpoint, the consequence of velocity-changing collisions in
buffered cells is a Brownian or diffusion motion of the atoms. The internal
atomic dipoles, \textit{e.g.}, those corresponding to the superposition
between the two Raman levels, diffuse across the variations of the light
fields. It is the near degeneracy of the Raman levels and the relatively large
Raman wavelength that make the coherent diffusion effectual. The spatial
effect is most clearly appreciated in light-storage experiments, in which the
relative amplitude of the Raman fields is imprinted onto the spatial field of
dipoles, which subsequently undergoes diffusion. The evolution becomes more
complicated in slow-light experiments, in which the propagation of polaritons
--- a combined excitation of light and atomic coherence --- is affected
simultaneously by optical diffraction and atomic diffusion.

This field of research is largely motivated by applications, namely,
high-precision measurements, especially with spatial multi-pixel resolution
\cite{RomalisNature2003}; multi-mode quantum memories \cite{PolzikPRA2010};
and spatial information processing, either classical or quantum
\cite{LettNature2009}. Atomic motion crucially affects the spectral and
spatial resolution, sensitivity, and coherence time of these applications.

In this Colloquium, we review the recent progress in the understanding of
motional effects in Raman processes. Spin-exchange among the active atoms and
with a polarizable buffer-gas \cite{HapperRMP1997} as well as pressure
broadening \cite{PeachAIP1981,Corey1984} are beyond the scope of the paper. We
emphasize mostly the regime of a dense inert buffer gas, in which the active
atoms undergo perfect diffusion in the medium, and employ the complementary
spectroscopic and spatial viewpoints. In doing so, we hope to illustrate the
underlying mechanisms and their consequences in hot atomic media as well as in
similar systems.

\section{Raman spectra of diffusing atoms\label{sec_motional}}

\subsection{The Doppler-Dicke transition}

The Doppler shift of a radiator moving at a velocity $\mathbf{v}$ is given by
$\omega_{\text{Doppler}}=\mathbf{vq}$. The spectrum exhibits side-bands at
$\pm\mathbf{vq}$, if the radiator is confined within two walls and
periodically flips its direction. When the direction flips are frequent,
spectral components at the original frequency, as well as higher-order
harmonics emerge. For very frequent collisions, the carrier prevails,
completely suppressing the Doppler effect. This narrowing phenomenon is named
after
\textcite{Dicke1953}%
. The distance between collisions $\Lambda$, with respect to the radiation
wavelength $\lambda$, determines the narrowing factor. A movie clip in the
Supplementary Material illustrates the Doppler-Dicke transition in the
acoustic spectrum of a moving emitter, obtained numerically by following
\textcite{Dicke1953}%
.

Doppler broadening in vapor originates from a picture of individual atoms
distributed among velocity groups and experiencing distinct Doppler shifts.
The Maxwell-Boltzmann distribution $F\left(  \mathbf{v}\right)  =(2\pi
v_{T}^{2})^{-3/2}e^{-v^{2}/(2v_{T}^{2})}$ results in an inhomogenous
broadening of%
\begin{equation}
\Gamma_{\text{Doppler}}=v_{T}|\mathbf{q|}, \label{eq_Gamma_dop}%
\end{equation}
where $v_{T}=\sqrt{k_{B}T/m}$ is the thermal velocity and $m$ the atomic mass
($\Gamma_{\text{Doppler}}$ refers to $1\sigma$).

\begin{figure}
[ptb]
\begin{center}
\includegraphics[
height=4.0381cm,
width=8.7784cm
]%
{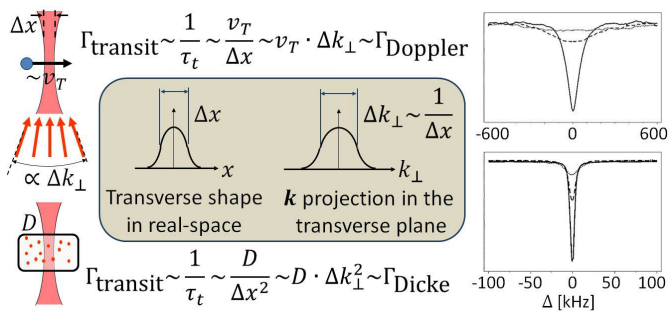}%
\caption{Transit-time interpretation of the Doppler and Dicke effects. A beam
of width $\Delta x$ has a span $\Delta k_{\bot}\sim1/\Delta x$ of transverse
momenta. The interaction time for an atom crossing the beam with velocity
$v_{T}$ (top) is $\tau_{t}=\Delta x/v_{T}$, resulting in a transit-time
broadening of $v_{T}\Delta k_{\bot}\sim\Gamma_{\text{Doppler}}$. For diffusing
atoms (bottom), the mean interaction time is $\Delta x^{2}/D$, leading to a
broadening of $D\Delta k_{\bot}^{2}\sim\Gamma_{\text{Dicke}}$. Right: Dark
resonances in rubidium vapor measured by \textcite{WeitzPRA2005} for a beam
diameter $\Delta x=5.6$ mm (top) without and (bottom) with $20$ Torr neon
buffer-gas. The respective linewidths are (solid lines) $2\Gamma
_{\text{Doppler}}=100$ kHz and $2\Gamma_{\text{Dicke}}=6$ kHz. Dashed and
dotted lines are measured with an angle between the Raman beams of
$\theta=0.31$ mrad and $\theta=0.62$ mrad, respectively. Adapted from
\textcite{WeitzPRA2005}.}%
\label{fig_TOF}%
\end{center}
\end{figure}

In a buffer-gas environment or due to confined cell geometries, the
velocity-groups picture breaks down, as collisions redistribute the velocities
faster than it takes the resonance to stabilize. Consequently, as we shall
establish in this section, the light merely faces fluctuations in the atomic
velocities, leading to a crossover from the Gaussian (inhomogenous) to a
Lorentzian (homogenous) lineshape. The average velocity associated with these
fluctuations is reduced with respect to $v_{T}$ by the Dicke narrowing factor:
$2\pi\Lambda/\lambda$. The homogenous Dicke half-width is thus given by
\cite{GalatryPR1961},%

\begin{equation}
\Gamma_{\text{Dicke}}\approx2\pi\frac{\Lambda}{\lambda}\Gamma_{\text{Doppler}%
}\ll\Gamma_{\text{Doppler}}. \label{eq_Gamma_dicke}%
\end{equation}
The Doppler effect corresponds to a ballistic motion of the atoms ($\Lambda
\gg\lambda$) and the Dicke effect to a diffusive motion ($\Lambda\ll\lambda$).
One finds that $\Gamma_{\text{Dicke}}$ is proportional to the diffusion
coefficient $D=v_{T}\Lambda$ and quadratic in the radiation wavenumber
\cite{Nelkin1964,Corey1984},%
\begin{equation}
\Gamma_{\text{Dicke}}=D\left\vert \mathbf{q}\right\vert ^{2}.
\end{equation}
Equations (\ref{eq_Gamma_dop}) and (\ref{eq_Gamma_dicke}) can intuitively be
understood as the inverse time an atom travels a distance $\lambda$
ballistically ($\propto\lambda/v_{T}$) or diffusively ($\propto\lambda^{2}%
/D$). Therefore, they are also interpreted as a transit-time broadening, as
illustrated in Fig. \ref{fig_TOF}. At low buffer-gas densities, when the
mean-free path is comparable to the wavelength ($\lambda/\Lambda\sim2\pi$),
the spectral width can be expressed as \cite{RautianSobelmanUSPEKHI1967}
\begin{equation}
\Gamma_{\text{Doppler-Dicke}}=\frac{v_{T}}{\Lambda}\frac{4}{a^{2}}H\left(
2\pi a\frac{\Lambda}{\lambda}\right)  , \label{eq_inter_width}%
\end{equation}
where $a^{2}=2/\ln2,$ and $H(x)=e^{-x}-1+x$ conveys at its limits the Doppler
trend [$H(x\rightarrow\infty)=x$] and the Dicke trend [$H(x\rightarrow
0)=x^{2}/2$].%

The condition $\Lambda\ll\lambda$ can hardly be satisfied for optical
resonances\ without introducing to much decoherence due to collisions. For
instance, room-temperature rubidium with $v_{T}\approx170$ m/s exhibits
$\Gamma_{\text{Doppler}}\approx220$ MHz at $\lambda=780$ nm. For this
wavelength, neon buffer-gas at a pressure of about $200$ Torr is required for
entering the Dicke regime $2\pi\Lambda/\lambda\sim1.$ At this pressure, the
decoherence induced by the neon on the optical resonance results in an
overwhelming pressure broadening of about 2 GHz \cite{GallagherPRA1974}.
Optical lines therefore remain Doppler broadened in nearly all thermal media.

For ground-state atomic transitions, buffer gases at the $1-100$ Torr levels
have been used since 1955 to delay the atomic motion and reduce Doppler and
transit-time broadening \cite{HapperRMP1972}. Since these transitions survive
millions of collisions with the buffer gas before decohering, and since the
associated microwave and rf wavelengths are much larger than the optical
wavelength, Dicke narrowing becomes far more reachable \cite{FrueholzJPB1985}.
As laid out in a pioneering work by
\textcite{Cyr1993}
and discussed in the rest of this section, all-optical Raman processes based
on these transitions were shown to exhibit roughly the same motional
broadening behavior, with the necessary adjustments due to the optical Doppler broadening.

\subsection{Motional broadening in Raman processes}

\begin{figure}
[ptb]
\begin{center}
\includegraphics[
height=4.1112cm,
width=7.9654cm
]%
{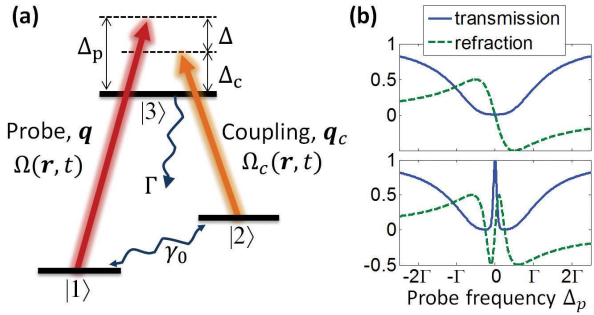}%
\caption{(color online) Electromagnetically-induced transparency in a
$\Lambda-$scheme. (a) The Raman resonance $\left\vert 1\right\rangle
\leftrightarrow\left\vert 2\right\rangle $ is excited via the state
$\left\vert 3\right\rangle $ by 'probe' and 'coupling' light fields. (b) Top:
transmission of the probe (solid line) in the absence of the coupling field
and accompanied refraction (dashed line). Bottom: dark resonance induced by
the coupling field.}%
\label{fig_EIT_scheme}%
\end{center}
\end{figure}

We consider as a model system dark resonances created via EIT in a $\Lambda
-$configuration, depicted in Fig. \ref{fig_EIT_scheme}(a). In $\Lambda-$EIT, a
$\emph{probe}$ field $\mathbf{E}$ and a \emph{coupling} field $\mathbf{E}_{c}$
couple two states from the atomic ground level ($\left\vert 1\right\rangle $
and $\left\vert 2\right\rangle $) to a common excited state ($\left\vert
3\right\rangle $). The fields are hereafter assumed to be classical and
characterized by the Rabi frequencies $\Omega$ and $\Omega_{c}$ via
$\mathbf{E}=\operatorname{Re}(\hbar\mathbf{\varepsilon}\Omega/\mu_{31})$ and
$\mathbf{E}_{c}=\operatorname{Re}(\hbar\mathbf{\varepsilon}_{c}\Omega_{c}%
/\mu_{32}),$ where $\mathbf{\varepsilon}$\textbf{, }$\mathbf{\varepsilon}_{c}%
$\ are the field polarizations and $\mu_{31}$, $\mu_{32}$ the transition
dipole moments. In the absence of the coupling field, the probe experiences
resonant absorption $\exp(-2\alpha L),$ determined by the absorption
coefficient $2\alpha$ and the medium length $L$. The combined action of the
probe and the coupling fields (the latter being usually much stronger,
$|\Omega_{c}|^{2}\gg|\Omega|^{2}$) drives the atoms into a dark state --- a
coherent superposition of the two lower states that inhibits the absorption of
the probe, rendering the medium transparent. One can easily verify that the
dark state on resonance $\Omega_{c}^{\ast}\left\vert 1\right\rangle
-\Omega^{\ast}\left\vert 2\right\rangle $ is decoupled from the excited state
$\left\vert 3\right\rangle $ under the influence of the interaction
Hamiltonian
\begin{equation}
H_{I}=-\hbar\Omega\left\vert 3\right\rangle \left\langle 1\right\vert
-\hbar\Omega_{c}\left\vert 3\right\rangle \left\langle 2\right\vert
+\text{h.c.}, \label{eq_HI}%
\end{equation}
essentially due to destructive interference between the two excitation paths
to $\left\vert 3\right\rangle $.

The dark resonance depends on the two-photon (Raman) detuning $\Delta
=\Delta_{p}-\Delta_{c}$ where $\Delta_{p}$ and $\Delta_{c}$ are, respectively,
the one-photon (optical) detunings of the probe and coupling fields, and
requires that $\Delta$ be smaller than the Raman linewidth. The latter varies
from Hz to tens of MHz in thermal vapor and is determined primarily by the
ground-state decoherence rate $\gamma_{0},$ power broadening from the coupling
light, and motional broadening. For comparison, in most cases, the optical
linewidth is much broader, varying from a few MHz for stationary (cold) atoms
to a few hundreds of MHz in Doppler-broadened systems. Therefore a narrow
transparency window forms at $\Delta_{p}=\Delta_{c}$ within the optical
absorption line \cite{HarrisPRL1991}, as can be seen in Fig.
\ref{fig_EIT_scheme}(b). At the same time, the probe also experiences very
steep dispersion $\omega(dn/d\omega)\gg1$ (dashed curve), leading to a much
reduced group-velocity. Ultra-narrow dark resonances are used in a wide
variety of processes, such as slow light \cite{HauNature1999}, stored light
\cite{LukinRMP2003}, and non-linear optics at low light levels
\cite{HarrisHauPRL1999,PeyronelNature2012}.%

The Raman detuning is sensitive to the difference between the Doppler shifts
of the probe and the coupling fields. When $\mathbf{q}=\mathbf{q}_{c}$, there
is no residual Doppler effect, and only the optical transitions are Doppler
broadened. In a general situation however, the Raman wavevector
\begin{equation}
\mathbf{k=q}-\mathbf{q}_{c} \label{eq_k_q1_q2}%
\end{equation}
does not vanish, and the expected residual widths are%
\begin{equation}
\Gamma_{\text{Doppler}}^{\text{res.}}=v_{T}k~~~~;~~~\Gamma_{\text{Dicke}%
}^{\text{res.}}=Dk^{2}~. \label{eq_residual_widths}%
\end{equation}
where $k=\left\vert \mathbf{k}\right\vert .$ The ratio between $\Gamma
_{\text{Dicke}}^{\text{res.}}$ and $\Gamma_{\text{Doppler}}^{\text{res.}}$,
the Dicke narrowing factor, ranges between $10^{-1}$ to $10^{-5}$ for typical
experimental conditions.

A chief example is the dark resonance among the two hyperfine sublevels of
ground-state alkali atoms, such as rubidium or cesium \cite{Akulshin1991}. The
hyperfine splitting, on the order of a few GHz, results in a Raman wavelength
$\lambda_{R}=2\pi/k$ on the order of a few centimeters for collinear beams,
implying a residual Doppler width of tens of kHz in the absence of a buffer
gas. With a typical buffer-gas pressure of $10$ Torr, the mean free-path of
the alkali atoms in the buffer gas is on the order of micrometers
(alkali-alkali collisions cause decoherence but are much more rare). The
narrowing factor is therefore on the order of $\Lambda/\lambda_{R}=10^{-4},$
eliminating completely the residual Doppler effect. A systematic measurement
of Dicke narrowing in dark resonances was reported by
\textcite{MeschedePRA1997}
for cesium (Fig. \ref{fig_mechede}), and later on by
\textcite{HelmPRA2000}
for rubidium, accompanied by a numerical model \cite{HelmPRA2001}\textsl{.}
The remaining homogenous width, due to alkali-alkali collisions, transit-time
broadening, wall collisions, and spin-destruction collisions with the buffer
gas, is on the order of tens of Hz, enabling the implementation of high
accuracy all-optical frequency standards
\cite{Cyr1993,WynandsPRA1999,knappe:1460}.%

\begin{figure}
[ptb]
\begin{center}
\includegraphics[
height=3.5353cm,
width=8.7784cm
]%
{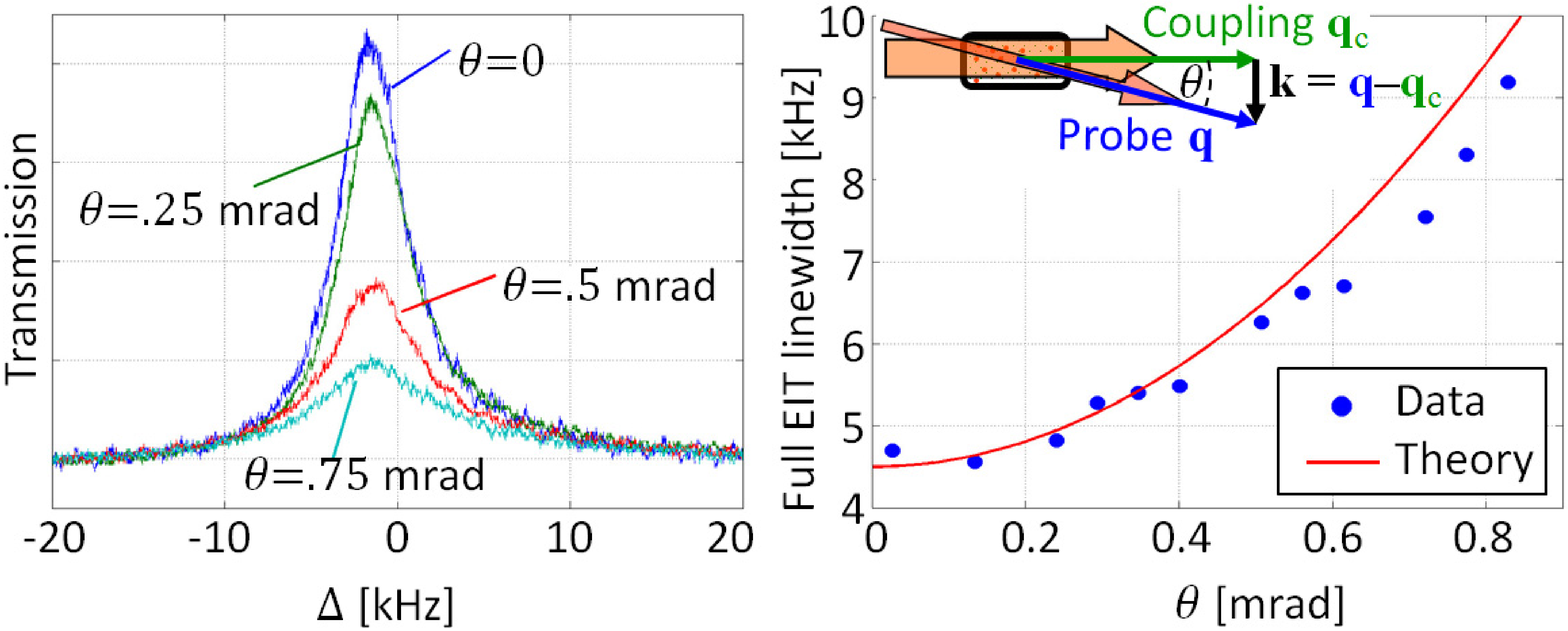}%
\caption{(color online) Residual Doppler-Dicke broadening of dark resonances
due to an angular deviation $\theta$ between the Raman beams, measured in Rb
vapor with $10$ Torr neon. The linewidth depends quadratically on $\theta,$ as
both the Doppler width ($\Gamma_{\text{Doppler}}^{\text{res.}}=\theta
\Gamma_{\text{Doppler}}^{\text{.}}$) and the Dicke narrowing factor
($2\pi\Lambda/\lambda_{R}=\theta\Lambda q$) are linear in $\theta$. The mean
free-path $\Lambda\approx2$ $\mu$m used in the theory (right, solid line)
corresponds to a collision rate of $\sim10^{8}/$sec and is calculated from the
Rb-Ne collisional cross-section \cite{GrafPRA1995,GallagherPRA1991}. The
broadening also leads to a decrease of the resonance transmission. Adapted
from \textcite{ShukerPRA2007}.}%
\label{fig_dicke_angle}%
\end{center}
\end{figure}

\textcite{Tabosa2004_angular_dependence}
measured the residual Doppler broadening in hyperfine dark-resonances by
introducing an angular deviation $\theta$ between the probe and the coupling
beams. Measurements of residual Dicke narrowing in buffered cells were
performed by
\textcite{WeitzPRA2005}
(Fig. \ref{fig_TOF}, right) and
\textcite{ShukerPRA2007}
(Fig. \ref{fig_dicke_angle}) in a degenerate $\Lambda-$scheme,\ using two
Zeeman states from the same hyperfine level so that $\left\vert \mathbf{q}%
\right\vert =\left\vert \mathbf{q}_{c}\right\vert .$ In this scheme,
$k=\left\vert \mathbf{q}-\mathbf{q}_{c}\right\vert \approx\theta\left\vert
\mathbf{q}\right\vert $ for small $\theta$, featuring $\lambda_{R}\approx1$ mm
for $\theta=1$ mrad. For a mean free-path of a few micrometers, one finds
$\lambda<\Lambda\ll\lambda_{R},$ \textit{i.e.,} the Raman resonance is in the
Dicke regime, while the optical resonance ($\lambda\lesssim1$ $\mu$m) is
Doppler broadened. The latter is virtually insensitive to $\theta$ and can be
as large as a few GHz, also due to pressure broadening. The $\theta$
dependence in Fig. \ref{fig_dicke_angle} exhibits the quadratic signature of
diffusion, with a clear narrowing effect: at $\theta=0.5$ mrad, the measured
width is $\Gamma_{\text{Dicke}}^{\text{res.}}=2$ kHz, much smaller than
$\Gamma_{\text{Doppler}}^{\text{res.}}=250$ kHz.%
\qquad

The light intensity has a strong effect on the Raman spectra, due to optical
pumping and the accompanying decoherence. The latter results in the so-called
power broadening of the natural width $\gamma_{0}$. For an atom at rest, the
optical pumping rate and the EIT power-broadening are given by $\gamma
_{p}=\left\vert \Omega_{c}\right\vert ^{2}/\Gamma.$ Both become smaller in a
Doppler broadened medium, because the effectiveness of the pumping varies
between the different velocity groups. This is a one-photon motional effect,
in which each velocity group experiences different pumping and decoherence
rates, providing inhomogenous 'conditions' for the Raman process. The
velocity-selective optical pumping \cite{AminoffJPhys1982,Gawlik1986} results
in correlations between the Raman and optical processes, similar to those
employed in the well-known techniques of Doppler-free saturated-absorption
spectroscopy \cite{SchawlowPRL1971} or laser-induced line narrowing
\cite{JavanPR1969}. Naturally, buffer gas and velocity-changing collisions
play an important role here, for example by allowing the cumulative optical
pumping of the whole Doppler profile or, alternatively, by limiting the
interaction time with a certain velocity group \cite{BjorkholmPRA1982}. These
correlations were studied for dark resonances\footnote{Even more intricate
correlations arise in Raman schemes involving two coupling fields, such as
4-wave mixing and electromagnetically-induced absorption. Here, the optical
dipoles, and not only the ground-state's populations and damping, become
velocity dependent \cite{TilchinPRA2011}.} in experiments by
\textcite{Zibrov2002_width_of_EIT}
and later by
\textcite{LvovskyOL2006}
and
\textcite{GoldfarbEPL2008}%
, along with theoretical analysis by
\textcite{JavanPRA2002}
and
\textcite{JavanAPB2003}%
. Being essentially a one-photon effect, it is beyond the scope of this
review; further details can be found in recent papers by
\textcite{XiaoMPLB2009}%
\ and
\textcite{GhoshPRA2009}%
, and in references therein.

\begin{figure}
[ptb]
\begin{center}
\includegraphics[
height=3.9185cm,
width=7.1171cm
]%
{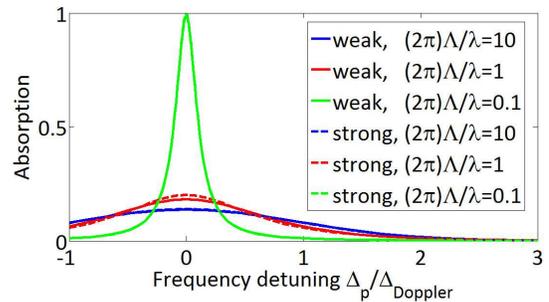}%
\caption{(color) Absorption spectra of thermal atoms. In the absence of
collisions, the homogenous linewidth $\Gamma$ is dominated by the Doppler
width $\Gamma_{\text{Doppler}}=100\Gamma$. The transition from the (blue)
Doppler limit to the (green) Dicke limit occurs when the effective mean
free-path $\Lambda=v_{T}/\gamma_{c}$ ($\gamma_{c}^{-1}$ is the velocity
correlation time due to collisions) is comparable to the wavelength $\lambda$
(red). The spectra for the (solid line) weak and (dashed line) strong
collisions are calculated from Eqs. (\ref{2l_s_final}) and (\ref{chi_1p}),
respectively. The differences between weak and strong collisions are not
distinguishable at the far limits (blue and green).}%
\label{fig_spectra_weak_strong}%
\end{center}
\end{figure}

In the absence of additional relaxation, the spectral line at the extreme
Doppler and Dicke limits is always, respectively, a Gaussian and a Lorentzian.
In the intermediate regime, however, it is determined by the nature of the
collisions. Depending mostly on the colliding species, the collisions may
either be strong (=hard) or weak (=soft), resulting in, respectively, a large
or small relative change in the velocity upon a single collision. A
phenomenological characterization of the collision strength is given by
\textcite{KeilsonStorer1952}
in their popular collision kernel. For a given collision rate, the kernel
renders the mean free path $\Lambda$ and the velocity correlation time
$\gamma_{c}^{-1}=\Lambda/v_{T}$. There is a vast literature dealing with the
sensitivity of atomic spectra to the nature of collisions, see
\textcite{BermanPRA1980,BloembergenPRA1984,SzudyPRA2001}%
, and references therein. Steady-state experiments, and spectroscopy in
particular, depend relatively weakly on the collision strength, as shown in
Fig. \ref{fig_spectra_weak_strong}. More elaborate schemes are required to
directly quantify the collision kernels, \textit{e.g.}, tagging of velocity
groups by selective optical-pumping in dilute buffer-gas and the subsequent
probing of the velocity redistribution
\cite{GallagherPRA1991,HapperPRA2010,HapperPRL2012}. An analogous problem with
trapped cold atoms undergoing elastic collisions was addressed by
\textcite{SagiPRL2010}%
.

Most of the work discussed in this Colloquium is carried out at the limits
$\lambda_{R}/\Lambda\gg2\pi$ or $\lambda_{R}/\Lambda\ll2\pi$, in which the
collision strength has negligible effect. In what follows, we shall
nevertheless introduce both approaches, \textit{i.e.},\textit{ }the Gaussian
process at the weak-collision limit and the Boltzmann relaxation at the
strong-collision limit, and show their equivalence in the far Doppler and
Dicke limits. A reader less interested in the mathematical derivation of the
spectra may proceed directly to subsection \ref{sec_trans_amp}.%

\subsubsection{Weak-collisions formalism}

We shall derive the Raman spectrum in the weak-collisions limit for stationary
uniform fields (plane waves), a weak probe, and no power broadening. Assuming
the $\Lambda-$atom of Fig. \ref{fig_EIT_scheme} travels along the classical
trajectory $\mathbf{r}=\mathbf{r}\left(  t\right)  ,$ either ballistic or
diffusive, we plug the time-dependent Rabi frequencies%
\begin{equation}
\tilde{\Omega}\left(  t\right)  =\Omega e^{i\mathbf{qr}\left(  t\right)
-i\omega t},\text{ }\tilde{\Omega}_{c}\left(  t\right)  =\Omega_{c}%
e^{i\mathbf{q}_{c}\mathbf{r}\left(  t\right)  -i\omega_{c}t} \label{Omega_n}%
\end{equation}
into the Hamiltonian (\ref{eq_HI}), with $\omega=c\left\vert \mathbf{q}%
\right\vert $ and $\omega_{c}=c\left\vert \mathbf{q}_{c}\right\vert $. To
account for relaxations, the individual atom is represented by a density
matrix $\rho_{ss^{\prime}}^{i}(t)$ ($s,s^{\prime}=1,2,3$) in a master equation
formalism, see for example,
\textcite{Cyr1993}
and
\textcite{NikonovQuantOpt1994}%
. For brevity, we shall characterize the relaxation of the optical dipole
($3\leftrightarrow1,2$) with a single decay rate $\Gamma,$ dominated by
pressure broadening. The ground-state relaxation rate is $\gamma_{0}.$ For a
given atomic density $n_{0},$ the absorption of the probe is calculated from
the imaginary part of the linear susceptibility\footnote{We define a linear
susceptibility $\chi,$ such that the transfer function of the probe field is
$\exp(i\chi z),$ as opposed to the prevailing (unitless) definition
$\exp(i|\mathbf{q}|\chi z/2)$ \cite{FleischhauerRMP2005}.} $\chi\left(
-\omega,\omega\right)  =g\langle\rho_{31}^{i}(t)/\tilde{\Omega}(t)\rangle$
where $g=\left\vert \mathbf{q}\right\vert n_{0}|\mu_{31}|^{2}/(\hbar
\epsilon_{0})$, $\epsilon_{0}$ is the vacuum permittivity, and $\langle
\rangle\equiv\lim_{\tau\rightarrow\infty}\int_{0}^{\tau}\frac{dt}{\tau}$.

We assume that the equilibrium state of the atom in the absence of the probe
is $\left\vert 1\right\rangle \left\langle 1\right\vert $ ($\rho
_{11}^{\text{eq.}}=1$), regardless of the velocity and the instantaneous
coupling power, which conforms with the limit of no power-broadening. The
first-order correction to the equilibrium state in the non-saturated and
weak-probe conditions $\Omega\ll\Omega_{c}\ll\Gamma$ involves only the
ground-state dipole $\rho_{21}^{i}(t)$ and the probe transition dipole
$\rho_{31}^{i}(t)$ \cite{KofmanPRA1997}:%
\begin{align}
\frac{d}{dt}\rho_{31}^{i}  &  =i\tilde{\Omega}_{c}\left(  t\right)  \rho
_{21}^{i}+i\tilde{\Omega}\left(  t\right)  \rho_{11}^{\text{eq.}}-i\left(
\omega-\Delta_{p}-i\Gamma\right)  \rho_{31}^{i},\nonumber\\
\frac{d}{dt}\rho_{21}^{i}  &  =i\tilde{\Omega}\,_{c}^{\ast}\left(  t\right)
\rho_{31}^{i}-i\left(  \omega-\omega_{c}-\Delta-i\gamma_{0}\right)  \rho
_{21}^{i}. \label{rho21_dot}%
\end{align}
To obtain $\rho_{31}^{i}(t),$ Eqs. (\ref{rho21_dot}) can be integrated and
solved formally, by iterations up to first order in $\rho_{31}^{i}$, a valid
approximation in the absence of power broadening $\left\vert \Omega_{c}%
^{2}\right\vert \ll\Gamma\gamma_{0}.$ In this regime, the susceptibility
becomes a sum $\chi\left(  -\omega,\omega\right)  =\chi_{\text{I}}%
(\omega)-\left\vert \Omega_{c}^{2}\right\vert \chi_{\text{II}}(\omega)$ of the
Raman resonance $\left\vert \Omega_{c}^{2}\right\vert \chi_{\text{II}}$ within
the optical resonance $\chi_{\text{I}}$ \cite{FirstenbergPRA2007}:%
\begin{subequations}
\begin{align}
\chi_{\text{I}}  &  =g\left\langle i%
{\displaystyle\int\limits_{0}^{t}}
dt_{1}e^{\left(  -i\Delta_{p}-\Gamma\right)  (t-t_{1})}e^{i\Phi_{\text{I}}%
}\right\rangle ,\label{eq_SI}\\
\chi_{\text{II}}  &  =g\left\langle i%
{\displaystyle\int\limits_{0}^{t}}
dt_{1}%
{\displaystyle\int\limits_{0}^{t_{1}}}
dt_{2}%
{\displaystyle\int\limits_{0}^{t_{2}}}
dt_{3}\frac{e^{\left(  -i\Delta_{p}-\Gamma_{1}\right)  \left(  t-t_{1}%
+t_{2}-t_{3}\right)  }}{e^{-\left(  i\Delta-\gamma_{0}\right)  \left(
t_{1}-t_{2}\right)  }}e^{i\Phi_{\text{II}}}\right\rangle . \label{eq_SII}%
\end{align}
The phases accumulated due to atomic motion though the light fields are
$\Phi_{\text{I}}=\mathbf{q\cdot}\left[  \mathbf{r}\left(  t\right)
-\mathbf{r}\left(  t_{1}\right)  \right]  $ and $\Phi_{\text{II}}%
=\mathbf{q}_{c}\mathbf{\cdot}\left[  \mathbf{r}\left(  t_{1}\right)
-\mathbf{r}\left(  t_{2}\right)  \right]  -\mathbf{q\cdot}\left[
\mathbf{r}\left(  t\right)  -\mathbf{r}\left(  t_{3}\right)  \right]  .$

At this point, one may recognize a homogenous Lorentzian line $%
{\textstyle\int}
d\tau e^{\left(  -i\Delta_{p}-\Gamma\right)  \tau}$ in Eq. (\ref{eq_SI}),
broadened by the motional phase $\mathbf{q\cdot r}\left(  \tau\right)  .$ This
is where the weak-collisions limit enters: As laid out by
\textcite{KuboBook1962}
and
\textcite{RautianSobelmanUSPEKHI1967}%
, the assumption of a Gaussian process for the random variable $\Phi
_{\text{I}}$, together with a Markovian velocity relaxation $\left\langle
\mathbf{\dot{r}}\left(  t\right)  \mathbf{\dot{r}}\left(  t-\tau\right)
\right\rangle =3v_{T}^{2}e^{-\gamma_{c}\left\vert \tau\right\vert },$ renders
the dephasing $\langle e^{i\Phi_{\text{I}}\left(  t,t-\tau\right)  }%
\rangle\approx e^{-\langle\Phi_{\text{I}}^{2}\rangle/2}\approx
e^{-|\mathbf{q|}^{2}\Lambda^{2}H\left(  \gamma_{c}\tau\right)  }$, with
$H(x)=e^{-x}-1+x$ and $\Lambda=v_{T}/\gamma_{c}$. This result leads to an
optical spectrum in the form of a Gumbel distribution \cite{GalatryPR1961}:
\end{subequations}
\begin{equation}
\chi_{\text{I}}(\Delta_{p})=ig\int_{0}^{\infty}d\tau e^{\left(  -i\Delta
_{p}-\Gamma\right)  \tau}e^{-|\mathbf{q}|^{2}\Lambda^{2}H\left(  \gamma
_{c}\tau\right)  }. \label{2l_s_final}%
\end{equation}
The absorption line $\operatorname{Im}\chi_{\text{I}}$ is shown in Fig.
\ref{fig_spectra_weak_strong}: At the Doppler limit $H(x)\approx x^{2}/2$
(solid blue), it is a Gaussian $\exp(-\Delta_{p}^{2}/\Gamma_{\text{Doppler}%
}^{2}/2)$; At the Dicke limit $H(x)\approx x$ (green), it is a Lorentzian
$[\Gamma+\Gamma_{\text{Dicke}}]/[\Delta^{2}+(\Gamma+\Gamma_{\text{Dicke}}%
)^{2}]$; and in between (red), it is neither.

A more elaborate but analogous derivation was performed by
\textcite{FirstenbergPRA2007}
for the Raman dephasing $\langle e^{i\Phi_{\text{II}}}\rangle$, resulting in a
closed integral form for $\chi_{\text{II}}.$ The Doppler-Dicke transition of
the Raman resonance was thereby formally obtained for the first time, for the
predominant case of a Doppler broadened optical line and a nearly resonant
coupling light:
\begin{equation}
\chi_{\text{II}}(\Delta)=\frac{ig}{\Gamma^{2}}\int_{0}^{\infty}d\tau
e^{\left(  i\Delta-\gamma_{0}\right)  \tau}e^{-k^{2}\Lambda^{2}H\left(
\gamma_{c}\tau\right)  }. \label{2l_sII}%
\end{equation}
Remarkably, the transmission line (\ref{2l_sII}) has the same form as the
absorption line (\ref{2l_s_final}), with the Raman parameters ($k,\gamma_{0}$)
replacing the optical parameters ($|\mathbf{q|},\Gamma$).

\subsubsection{Strong-collisions formalism}

For the strong-collisions formalism, we shall use a density-matrix
distribution function in space and velocity $\tilde{\varrho}_{ss^{\prime}%
}=\tilde{\varrho}_{ss^{\prime}}(\mathbf{r},\mathbf{v},t),$ constructed from
the sum over (identical) individual atoms:%
\begin{equation}
\tilde{\varrho}_{ss^{\prime}}=\sum_{i}\rho_{ss^{\prime}}^{i}\left(  t\right)
\delta\left(  \mathbf{r}-\mathbf{r}_{i}\left(  t\right)  \right)
\delta\left(  \mathbf{v}-\mathbf{v}_{i}\left(  t\right)  \right)  .
\end{equation}
This approach, first used by
\textcite{May1999}
in this context, is general in that it allows atoms in different states to
travel or diffuse between the illuminated and the dark areas, both in the real
spatial space and in velocity space, and thereby circumvents the approximation
of an open system \cite{NikonovQuantOpt1994}. In a hot vapor, the
density-matrix distribution can be taken as classical in the external-motion
degrees of freedom, and evolves according to
\begin{align}
&  \left(  \partial_{t}+\mathbf{v\cdot}\partial_{\mathbf{r}}\right)
\tilde{\varrho}_{ss^{\prime}}+\left(  \partial_{t}\tilde{\varrho}_{ss^{\prime
}}\right)  _{\operatorname{col}.}\label{eq_rhoss}\\
&  =\sum_{i}\left(  \partial_{t}\rho_{ss^{\prime}}^{i}\right)  \delta\left(
\mathbf{r}-\mathbf{r}_{i}\left(  t\right)  \right)  \delta\left(
\mathbf{v}-\mathbf{v}_{i}\left(  t\right)  \right)  ,\nonumber
\end{align}
where $\left(  \partial_{t}\tilde{\varrho}_{ss^{\prime}}\right)
_{\operatorname{col}.}$ accounts for collisions. The right-hand side of Eq.
(\ref{eq_rhoss}) describes the internal atomic dynamics, which can be taken
from Eqs. (\ref{rho21_dot}). Here however, to set the stage for the
description of polariton dynamics, let us generalize Eqs. (\ref{rho21_dot})
and employ a structured (time-dependent) probe and a structured (stationary)
coupling:%
\begin{equation}
\tilde{\Omega}=\Omega\left(  \mathbf{r},t\right)  e^{i\mathbf{qr}-i\omega
t},~~~\tilde{\Omega}_{c}=\Omega_{c}\left(  \mathbf{r}\right)  e^{i\mathbf{q}%
_{c}\mathbf{r}-i\omega_{c}t}, \label{eq_rabi_envelopes}%
\end{equation}
where $\Omega\left(  \mathbf{r},t\right)  $ and $\Omega_{c}\left(
\mathbf{r}\right)  $\ are slowly-varying envelopes of the Rabi-frequencies.
Correspondingly, we define the slowly-varying atomic densities $\varrho_{31}=$
$\tilde{\varrho}_{31}e^{i\omega t-i\mathbf{qr}}$ and $\varrho_{21}%
=\tilde{\varrho}_{21}e^{i\left(  \omega-\omega_{c}\right)  t-i\left(
\mathbf{q}-\mathbf{q}_{c}\right)  \cdot\mathbf{r}}$.

We shall now consider the strong-collisions limit. In this limit, a single
collision is enough to completely randomize the atomic velocity. Here we
assume that the post-collision velocity is drawn from the equilibrium
distribution $F\left(  \mathbf{v}\right)  ,$ regardless of the pre-collision
velocity; the generalization to velocity-dependent kernels can be performed
along the same lines \cite{MayPRA2001,GhoshPRA2009}. These assumptions pertain
to a Kubo-Anderson process, which in principle could be implemented in the
individual-atom formalism used above for the weak-collisions limit
\cite{BrissaudFrischJMathPhys1974,SagiPRL2010}. In practice however,
calculating the four-time dephasing of the Raman resonance [$\Phi_{\text{II}}$
in Eq. (\ref{eq_SII})] under the Kubo-Anderson assumptions is prohibitive. We
thus resort to a more direct approach and invoke a Boltzmann collision term
with a single relaxation rate $\gamma_{c}$ \cite{Nelkin1964}:%
\begin{equation}
\left(  \partial_{t}\tilde{\varrho}_{ss^{\prime}}\right)  _{\operatorname{col}%
.}=-\gamma_{c}\left[  \varrho_{ss^{\prime}}(\mathbf{r},\mathbf{v}%
,t)-\rho_{ss^{\prime}}(\mathbf{r},t)F\left(  \mathbf{v}\right)  \right]  ,
\label{rs-5}%
\end{equation}
where the spatial density-matrix is%
\begin{equation}
\rho_{ss^{\prime}}(\mathbf{r},t)=\int d^{3}v\varrho_{ss^{\prime}}%
(\mathbf{r},\mathbf{v},t). \label{rs-4}%
\end{equation}
The physical meaning of $\rho_{ss^{\prime}}(\mathbf{r},t)$ is readily
understood by identifying its diagonal elements $\rho_{ss}(\mathbf{r},t)$ as
the spatial density of atoms at state $\left\vert s\right\rangle ,$ and its
off-diagonal elements as the polarization density $\mathbf{P}\left(
\mathbf{r},t\right)  $, \textit{e.g.}, $\rho_{31}\left(  \mathbf{r},t\right)
=\mathbf{\varepsilon P}_{31}\left(  \mathbf{r},t\right)  /\mu_{31}^{\ast}.$
Note that Eq. (\ref{rs-5}) does not consider pressure broadening, which we
later introduce via the atomic decay rates \cite{Corey1984}.

Finally, identifying $\rho_{11}^{\text{eq.}}\Rightarrow n_{0}F(\mathbf{v})$ in
Eq. (\ref{rho21_dot}) and substituting the definitions
(\ref{eq_rabi_envelopes})-(\ref{rs-4}) in Eq. (\ref{eq_rhoss}), we obtain the
equations of motion for the densities:
\begin{subequations}
\label{ul-rl-4}%
\begin{gather}
\left[  \partial_{t}+\mathbf{v\cdot}\partial_{\mathbf{r}}-i\delta_{p}\left(
\mathbf{v}\right)  \right]  \varrho_{31}(\mathbf{r},\mathbf{v},t)-i\Omega
_{c}(\mathbf{r})\varrho_{21}(\mathbf{r},\mathbf{v},t)\nonumber\\
=\gamma_{c}\rho_{31}(\mathbf{r},t)F\left(  \mathbf{v}\right)  +i\Omega
(\mathbf{r},t)n_{0}F\left(  \mathbf{v}\right)  ,\\
\left[  \partial_{t}+\mathbf{v\cdot}\partial_{\mathbf{r}}-i\delta\left(
\mathbf{v}\right)  \right]  \varrho_{21}(\mathbf{r},\mathbf{v},t)-i\Omega
_{c}^{\ast}(\mathbf{r})\varrho_{31}(\mathbf{r},\mathbf{v},t)\nonumber\\
=\gamma_{c}\rho_{21}(\mathbf{r},t)F\left(  \mathbf{v}\right)  ,
\end{gather}
where $\delta_{p}\left(  \mathbf{v}\right)  =\Delta_{p}-\mathbf{q\cdot
v}+i\left(  \Gamma+\gamma_{c}\right)  $ and $\delta\left(  \mathbf{v}\right)
=\Delta-\left(  \mathbf{q}-\mathbf{q}_{c}\right)  \mathbf{\cdot v}+i\left(
\gamma_{0}+\gamma_{c}\right)  $ are the Doppler-shifted complex detunings.
These equations, together with a wave equation for the probe field, form the
basis for the diffusion of polaritons presented in the next section.

To derive the Doppler-Dicke profiles at this stage, we restrict Eq.
(\ref{ul-rl-4}) to stationary plane waves,
\end{subequations}
\begin{subequations}
\label{eq_unifrom_coherences}%
\begin{align}
i\delta_{1}(\mathbf{v})\varrho_{31}(\mathbf{v})+i\Omega_{c}\varrho
_{21}(\mathbf{v})  &  =-\left(  \gamma_{c}\rho_{31}+i\Omega n_{0}\right)
F(\mathbf{v}),\\
i\delta(\mathbf{v})\varrho_{21}(\mathbf{v})+i\Omega_{c}^{\ast}\varrho
_{31}(\mathbf{v})  &  =-\gamma_{c}\rho_{21}F(\mathbf{v}).
\end{align}
From Eqs. (\ref{eq_unifrom_coherences}),
\textcite{FirstenbergPRA2008}
derived an exact integral form for the susceptibility\footnote{The steady
state dipoles $\rho_{31}$ and $\rho_{21}$ are derived by formally solving Eqs.
(\ref{eq_unifrom_coherences}) for $\varrho_{31}\left(  \mathbf{v}\right)  $
and $\varrho_{21}(\mathbf{v})$, and integrating over velocities. The resulting
susceptibility is $\chi=ig\gamma_{c}^{-1}[(1-i\gamma_{c}G_{\delta_{p}})/G-1],$
where $G=(1-i\gamma_{c}G_{\delta_{p}})(1-i\gamma_{c}G_{\delta})+\gamma_{c}%
^{2}G_{|\Omega_{c}|}^{2}$ and $G_{X}=\int d^{3}vX(\mathbf{v})F\left(
\mathbf{v}\right)  /[\delta_{p}(\mathbf{v})\delta(\mathbf{v})-|\Omega_{c}%
|^{2}]$. Here, $X$ stands for either $\delta_{p}(\mathbf{v}),$ $\delta
(\mathbf{v}),$ or $\left\vert \Omega_{c}\right\vert $. Generally, a
calculation of the $G_{X}$'s integrals is required to obtain the concurrent
motional-broadening of the optical and dark resonances.} $\chi=(g/n_{0}%
)\rho_{31}/\Omega$ and exemplified numerically the Doppler-Dicke transition of
the dark resonance. The transition is similar to but not exactly as that found
in the weak-collisions limit. For the sake of elucidation, we may (as before)
examine the one-photon spectrum by substituting $\Omega_{c}=0$,
\end{subequations}
\begin{equation}
\chi_{\text{I}}=g\frac{G_{\text{I}}\left(  \Delta_{p}\right)  }{i\gamma
_{c}G_{\text{I}}\left(  \Delta_{p}\right)  -1}, \label{chi_1p}%
\end{equation}
where $G_{\text{I}}\left(  \Delta_{p}\right)  $ is the widely used Voigt
profile:
\begin{equation}
G_{\text{I}}\left(  \Delta_{p}\right)  =\frac{1}{\sqrt{2\pi}v_{T}}\int
du\frac{e^{-u^{2}/(2v_{T}^{2})}}{\Delta_{p}-|\mathbf{q}|u+i\left(
\Gamma+\gamma_{c}\right)  }. \label{g1}%
\end{equation}
The spectrum in the form of Eq. (\ref{chi_1p}) exhibits the Doppler-Dicke
transition; see discussion by
\textcite{May1999}
and references therein. A comparison in Fig. \ref{fig_spectra_weak_strong} to
the weak-collisions spectra reveals a maximal deviation of $10-20$ percent at
the Doppler-Dicke crossover.

\subsection{The Raman resonance at the diffusion limit\label{sec_trans_amp}}

We conclude this section by discussing the Raman lineshape at the Dicke limit,
for nearly degenerate, nearly collinear beams, such that $k=\left\vert
\mathbf{q}-\mathbf{q}_{c}\right\vert \ll\left\vert \mathbf{q}\right\vert .$ In
the vicinity of the Raman line ($\Delta_{p}\approx\Delta_{c}$), a closed set
of equations was obtained by
\textcite{FirstenbergPRA2008}%
\footnote{Briefly, Eq. (\ref{eq_rho31}) is obtained by integrating Eq.
(\ref{ul-rl-4}a) and solving for $\rho_{31}$, assuming it does not depend on
the non-equilibrium velocity-distribution of $\rho_{21}$. Taking $\gamma_{c}$
as the dominant rate in the ground-state dynamics, Eq. (\ref{ul-rl-4}b) is
integrated over velocity to obtain a continuity equation and a diffusive-flux
equation (Fick's first law) in terms of the current densities $\mathbf{J}%
_{ss^{\prime}}(\mathbf{r},t)=\int d^{3}v\mathbf{v}\varrho_{ss^{\prime}%
}(\mathbf{r},\mathbf{v},t).$ Finally, the condition $k\ll|\mathbf{q}|$ yields
Eq. (\ref{eq_rho21}).} for the optical dipoles $\rho_{31}\left(
\mathbf{r},t\right)  $:
\begin{equation}
\rho_{31}(\mathbf{r},t)=\frac{i}{\Gamma^{\prime}}\left[  \Omega(\mathbf{r}%
,t)n_{0}+\Omega_{c}(\mathbf{r})\rho_{21}(\mathbf{r},t)\right]  ,
\label{eq_rho31}%
\end{equation}
and for the ground-state dipoles $\rho_{21}\left(  \mathbf{r},t\right)  $:%
\begin{equation}
\left[  \partial_{t}-i\Delta+\gamma_{0}+\gamma_{P}(\mathbf{r}%
)-D(\mathbf{\nabla}+i\mathbf{k})^{2}\right]  \rho_{21}(\mathbf{r}%
,t)=S(\mathbf{r},t). \label{eq_rho21}%
\end{equation}
The ground-state dipoles obey a diffusion-like equation with the coefficient
$D=v_{T}^{2}/\gamma_{c}$ ($\mathbf{\nabla\equiv}\partial_{\mathbf{r}}$ is the
gradient). Here, $S(\mathbf{r},t)=-n_{0}\Omega_{c}^{\ast}(\mathbf{r}%
)\Omega(\mathbf{r},t)/\Gamma^{\prime}$ is a source term --- the effective
two-photon drive of the Raman resonance. $\gamma_{P}(\mathbf{r})=|\Omega
_{c}(\mathbf{r})|^{2}/\Gamma^{\prime}$ is a spatially varying power-broadening
rate. $\Gamma^{\prime}=\Gamma^{\prime}(\Delta_{p})\equiv ig/\chi_{\text{I}}$
is the one-photon (Voigt) spectrum from Eq. (\ref{chi_1p}) ($\Gamma^{\prime
}=\Gamma+i\Delta_{p}$ for stationary atoms). Notably, atomic motion affects
the Raman resonance both directly, due to dephasing of the Raman line, and
indirectly via the power broadening.

It is important to realize that the diffusion term $D\nabla^{2}$ in Eq.
(\ref{eq_rho21}) corresponds to the actual diffusion of the active atoms in
the buffered cell. In fact, the description of spatial diffusion of the
internal states of atoms and molecules in the form of Eq. (\ref{eq_rho21})
dates back to the seminal work by
\textcite{TorreyPR1956}
and has been the common practice for optical-pumping experiments in buffered
cells \cite{HapperRMP1972,ZmbonNuovoCi1980}. Accordingly, the term $Dk^{2}$
(for non-structured stationary beams $\partial_{t}=\mathbf{\nabla}=0$)
accounts for the diffusion of atoms across the fields' interference pattern.
The linear susceptibility is then easily obtained from Eqs. (\ref{eq_rho31})
and (\ref{eq_rho21}):%

\begin{equation}
\chi=\frac{g}{n_{0}}\frac{\rho_{31}}{\Omega}=\frac{ig}{\Gamma^{\prime}}\left(
1-\frac{\gamma_{P}}{\gamma+Dk^{2}-i\Delta}\right)  . \label{chi_bar_dicke}%
\end{equation}
The two terms in the brackets correspond to the optical resonance and to the
dark resonance. The latter is given as a complex Lorentzian, and its width is
the sum of the linewidth for stationary atoms $\gamma\equiv\gamma_{0}%
+\gamma_{P}$ and the motional broadening $Dk^{2}$ (Fig. \ref{fig_dicke_angle}).

\section{Polaritons Dynamics in Diffusive Media}

We have so far discussed the response of the atomic medium to a given
arrangement of light beams from a spectroscopic viewpoint, but have not
considered the spatial consequences of atomic motion. As these were taken into
account in the dynamic description of the density-matrix distributions, we may
now directly apply the results of the previous section to the evolution of the
structured light fields in space and time. The non-local response arising from
the atomic motion and reflected in the dependence of the linear susceptibility
on the wavevector has been demonstrated in recent years through various
processes and, in particular, with slow light. In principle, it is the
effective delay of the light in the form of a light-matter polariton
\cite{FleischhauerPRL2000}, becoming comparable to the atomic motion through
the beams, that renders these effects pronounced. That said, the description
of the phenomena reviewed in this section is not always an obvious spatial
consequence of atomic motion, and it is sometimes necessary to return to and
employ the spectral picture of a manifold of Doppler-Dicke spectra.

\begin{figure}
[ptb]
\begin{center}
\includegraphics[
height=3.9451cm,
width=7.9654cm
]%
{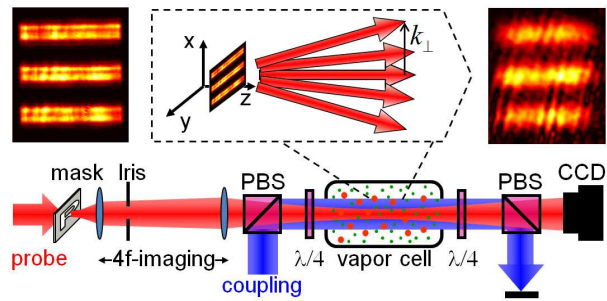}%
\caption{(Color online) Structured slow light. The probe beam is patterned by
using a transmission mask, a grating, or a spatial light-modulator and imaged
onto the cell. An iris may be used to filter high-frequencies and limit the
angular span. After the cell, the probe is imaged onto a camera. Top photos:
Diffusion of (left) a line pattern with 1.5 lines-pair/mm after (right) 6
$\mu$m delay. Taken in the setup of \textcite{ShukerImaging2007}.}%
\label{fig_diffusion}%
\end{center}
\end{figure}

Slow light structured in the plane normal to the propagation direction,
denoted as \emph{slow images}, exhibits remarkable properties. A notable
example is the delay and preservation of spatial quantum coherence and
entanglement, demonstrated by
\textcite{LettNature2009}%
. The 'image' may be complex, having both amplitude and phase patterns,
conforming to the amplitude and phase of the polariton's dark state. A typical
setup for a slow-image experiment is shown in Fig. \ref{fig_diffusion}: while
the coupling beam is large and uniform, the probe is patterned, imaged onto
the cell, and eventually recorded. If the probe is also temporally modulated
into a pulse, the pulse, and thus the whole image, is delayed in the medium.

The reduced group velocity of the probe follows directly from the linear
susceptibility, $v_{g}^{-1}=[d(\operatorname{Re}\chi)/d\Delta]_{\Delta=0}$ for
$v_{g}\ll c$. At the diffusion limit, for nearly resonant light ($\Delta
_{c}\approx\Delta_{p}\approx0$, for which the damping rates $\Gamma^{\prime}$
and $\gamma_{P}$ are real), Eq. (\ref{chi_bar_dicke}) gives $v_{g}%
=(\gamma+Dk^{2})^{2}/(\alpha\gamma_{P}).$ Here, $2\alpha=2g/\Gamma^{\prime}$
is the absorption coefficient with no coupling field. As also shown by
\textcite{ScullyPRL1999}
for the bufferless case (Doppler-broadened dark resonance), the group velocity
is $k-$dependent and only reverts to the known expression%
\phantomsection\label{eq_vg}
$v_{g}=\gamma^{2}/(\alpha\gamma_{P})$ for small enough $k$. For typical values
in hot vapor $\gamma\approx\gamma_{P}\approx10^{1}-10^{6}$ Hz and
$\alpha\approx1/$cm, $v_{g}$ is on the order of m/s to km/s
\cite{BudkerPRL1999}. The group delay in a medium of length $L\approx1-10$ cm
is then $\tau_{d}=L/v_{g}\approx1-10^{5}$ $\mu$s, easily comparable to the
time at which atoms can travel through the beam, or through the $0.1-10$ mm
features of an image, in both buffered and bufferless cells. In contrast, slow
images with $0.1$-mm feature size delayed for only 10 ns using optical (not
Raman) resonances by
\textcite{Howell2007_slowing_images}%
, showed no significant motional effects.%

\subsection{Transverse spreading of light}

The reflection of atomic motion in the spatial variation of slow light was
preceded by extensive research on general spatial consequences of EIT and
similar Raman processes. The emphasis, in the first experiments by
\textcite{KasapiPRL1995}%
,\
\textcite{MoseleyPRL1995}%
, and
\textcite{HarrisPRL1995b}%
, and in following years, was given to the implications of finite and
inhomogenous strong beams, inducing inhomogenous absorption and refraction,
and to the related effects of self-focusing and waveguiding. A direct
observation of slow-light spreading due to atomic motion was reported by
\textcite{PugatchPRL2007}%
, using a probe beam with a darkened (blocked) center. Images of the beam
taken on and off resonance showed that the $50$ $\mu$s slowing delay was
enough for the atomic diffusion in the buffered cell to 'fill' the
$0.5$-mm-diameter dark center almost completely. In effect, the ground-state
dipoles diffusing to the center stimulate the conversion of coupling light
into probe light. The phase pattern of the dipoles ensemble, originating from
the incoming probe and coupling fields, acts as a directional source for this
stimulated emission. In the alternative picture of polariton propagation, the
filling of the center is interpreted as diffusion of the polaritons due to
their atomic constituent.

\begin{figure}
[ptb]
\begin{center}
\includegraphics[
height=4.5542cm,
width=8.074cm
]%
{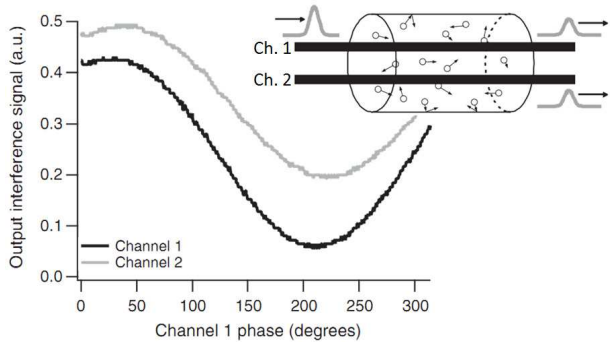}%
\caption{Phase coherence between the two 'output channels' of a slow-light
beam-splitter by \textcite{XiaoPRL2008}. The atoms moving in the bufferless
wall-coated cell mediate the coherence between the two channels. The measured
transfer efficiency ($<5\%$) is a function of the slowing delay ($0.5$ ms),
the decoherence due to wall collisions, and the beams/cell geometry.
Optimization of these factors promises efficiencies close to unity. Adapted
from \textcite{XiaoPRL2008}.}%
\label{fig_xiao2008}%
\end{center}
\end{figure}

A direct phase measurement of spreading light was reported by
\textcite{XiaoPRL2008}%
. Here the ballistic atomic motion in a bufferless, wall-coated cell is used
to coherently transfer light between adjacent optical modes, as shown in Fig.
\ref{fig_xiao2008}. Atomic coherence is created along the input channel and is
maintained as the atoms spread in the cell and collide with the walls. While
longitudinal spreading has no significance for the degenerate arrangement used
($\mathbf{q}=\mathbf{q}_{c}$), the transverse spreading stimulates the
coherent excitation of a propagating pulse in the second channel.%

\begin{figure}
[ptb]
\begin{center}
\includegraphics[
height=3.5131cm,
width=8.7761cm
]%
{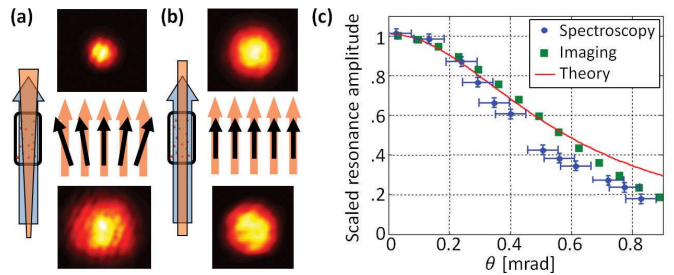}%
\caption{(Color online) Resonant transmission versus the angular deviation
$\theta$ between the Raman beams. (a) Imaging experiment, measuring the
resonant transmission of a diverging probe beam with a large (plane-wave)
coupling beam. Pictures are taken beyond the cell by impinging the beams
directly onto a CCD detector, (top) with and (bottom) without the atoms.
Stronger absorption is observed away from the center, where, in the optical
ray approximation, $\theta$ is larger. Maximal $\theta$ is $\sim2$ mrad at the
beam radius. (b) A reference with a non-diverging probe. (c) Scaled
transmission ($\square$) from the aforementioned imaging experiment and
($\bigcirc$) from spectroscopy (Fig. \ref{fig_dicke_angle}). The theoretical
curve is calculated from Eq. (\ref{EIT_Lorentzian_absorption}). Adapted from
\textcite{ShukerPRA2007}.}%
\label{fig_dicke_amplitude}%
\end{center}
\end{figure}

To understand the spreading of light within the Doppler-Dicke context, we
return to Eq. (\ref{chi_bar_dicke}) with nearly resonant beams ($\Delta
_{p}\approx\Delta_{c}\approx0$), for which the absorption spectrum of the
probe is given by
\begin{equation}
\operatorname{Im}\chi=\alpha\left[  1-\frac{\gamma_{P}(\gamma+Dk^{2})}%
{(\gamma+Dk^{2})^{2}+\Delta^{2}}\right]  . \label{EIT_Lorentzian_absorption}%
\end{equation}
The relative height at the center of the dark-resonance ($\Delta=0$) depends
on the Raman wavenumber $k$ in the form of a Lorentzian $\gamma_{P}%
/(\gamma+Dk^{2})$ of width $k_{0}=(\gamma/D)^{1/2},$ as confirmed by
\textcite{WeitzPRA2005}
and
\textcite{ShukerPRA2007}
with a small deviation angle $\theta\approx k/|\mathbf{q}|$ between the
coupling and probe beams (Fig. \ref{fig_dicke_amplitude}). The dependency of
the transmission on $\theta$ is manifested in experiments with non-uniform,
structured, light fields, due to the angular span of beam. In the
decomposition of the field into a manifold of superimposed plane waves,
high-order transverse modes and finely-patterned beams require a large angular
span, which implies large Raman wavenumbers (top sketch in Fig.
\ref{fig_diffusion}). When these are attenuated due to motional broadening,
the fine structure of the beam deteriorates. A maximum acceptance angle
$\theta=k_{0}/|\mathbf{q}|$ thus sets a minimum 'pixel' size of $2\pi/k_{0}$
that can be efficiently transmitted, whereas smaller features are bound to
spread. So atomic motion, via motional broadening, results in the spreading of
the light field.%

\subsection{Diffusion and motional-induced diffraction}

At certain conditions, motional broadening results in an exact diffusion of
the slow polaritons, as well as in a diffraction-like evolution. To this end,
we employ the following arrangement: a plane-wave coupling field along the
$z-$axis; a paraxial, nearly parallel, probe $\mathbf{q}\parallel
\mathbf{q}_{c}$ with a finite envelope in the transverse plane $\left(
x,y\right)  $; and a nearly degenerate Raman scheme $|\mathbf{q}_{c}|\approx$
$|\mathbf{q}|\equiv q$ so that the Raman wavevectors resulting from the
probe's structure have a negligible $z-$component (hyperfine splitting with
$\lambda_{R}$ on the order of centimeters is still permitted). Note that the
choice $\mathbf{q}=\mathbf{q}_{c}$ still allows for a small angular deviation
between the beams via a phase term $e^{i\theta qx}$ in the probe's envelope.
In the paraxial approximation, the probe field obeys%
\begin{equation}
\left(  \frac{\partial_{t}}{c}+\partial_{z}-i\frac{\nabla_{\perp}^{2}}%
{2q}\right)  \Omega\left(  \mathbf{r},t\right)  =i\frac{g}{n_{0}}\rho
_{31}\left(  \mathbf{r},t\right)  , \label{Ert-4}%
\end{equation}
where $\nabla_{\perp}^{2}=\partial_{x}^{2}+\partial_{y}^{2}$ is the transverse
Laplacian. Eqs. (\ref{eq_rho31}), (\ref{eq_rho21}), and (\ref{Ert-4}) compose
the full set of equations of motion for the slowly varying envelopes.

The group velocity $v_{g}$ obtained on page \pageref{eq_vg} is applicable for
pulses long enough such that their bandwidth (in the temporal frequency
domain) is within the linear dispersion regime. If the pulses vary more slowly
than any other rate in the system, the time dependence can be treated
parametrically, based on a quasi-steady-state assumption. The steady-state
assumption can easily be lifted within the linear response approximation,
which is valid as long as the coupling field is stationary and uniform.
Keeping in mind that the traveling pulses are essentially delayed, it will
still be meaningful in quasi-steady-state to translate distance to time via
$v_{g}$.

The changes of the probe along $z$ are due to its finite extent (in a pulsed
experiment) and due to absorption and refraction in the medium; both are
assumed to vary much more slowly than the envelope in the transverse plane,
making the diffusion negligible in the $z-$direction. The relevant Raman
wavevectors are thus identical with the transverse spatial frequencies. Taking
the Fourier transform $\left(  x,y\right)  \rightarrow(k_{x},k_{y}%
)=\mathbf{k}_{\perp}$ of Eqs. (\ref{eq_rho31}), (\ref{eq_rho21}), and
(\ref{Ert-4}) while maintaining the explicit $z$ dependence, one recovers the
linear susceptibility $\chi\left(  \mathbf{k}_{\perp}\right)  =i\alpha
\lbrack1-\gamma_{P}/(\gamma+Dk_{\perp}^{2}-i\Delta)]$ and the steady-state
evolution along $z$:
\begin{equation}
\partial_{z}\Omega(\mathbf{k}_{\perp};z)=\left[  i\chi\left(  \mathbf{k}%
_{\perp}\right)  -i\frac{k_{\perp}^{2}}{2q}\right]  \Omega(\mathbf{k}_{\perp
};z). \label{prop1}%
\end{equation}
Clearly, the geometric effect of free-space diffraction $ik_{\perp}^{2}/(2q)$
influences slow images precisely as in free space. For a confined $k_{\perp}%
-$spectrum, the susceptibility can be expanded in orders of $k_{\perp}^{2}$ as
$\chi\left(  \mathbf{k}_{\perp}\right)  =\chi_{0}+\left[  \chi\left(
\mathbf{k}_{\perp}\right)  \right]  _{\text{motional}},$ where $\chi
_{0}=i\alpha\lbrack1-\gamma_{P}/(\gamma-i\Delta)]$ is the susceptibility for
an atom at rest, and%
\begin{equation}
iv_{g}\left[  \chi\left(  \mathbf{k}_{\perp}\right)  \right]
_{\text{motional}}=\frac{-\gamma^{2}}{\left(  \gamma-i\Delta\right)  ^{2}%
}Dk_{\perp}^{2}+O(k_{\perp}^{4}), \label{eq_k2_k4}%
\end{equation}
with $v_{g}=\gamma^{2}/(\alpha\gamma_{P}).$ The $k_{\perp}^{4}$ term is
negligible when the probe's spectrum is initially confined within $k_{\perp
}\ll k_{0}=(\gamma/D)^{1/2}.$ The requirement $k_{\perp}\ll k_{0}$ is usually
stricter than the optical paraxial condition (for example, the typical values
$D=10$ cm$^{2}$/s and $\gamma=10$ kHz give $k_{0}$ on the order of $0.01$
$\mu$m$^{-1}$). Returning to $\left(  x,y\right)  $ space,
\begin{equation}
\partial_{z}\Omega=\left[  i\chi_{0}+\left(  \frac{\mathcal{D}}{v_{g}}%
+\frac{i}{2q}\right)  \nabla_{\perp}^{2}+O(\nabla_{\perp}^{4})\right]  \Omega,
\label{eq_effective_diffusion}%
\end{equation}
we find an effective complex diffusion coefficient:%
\begin{equation}
\mathcal{D}=D\frac{1-(\Delta/\gamma)^{2}}{\left[  1+(\Delta/\gamma
)^{2}\right]  ^{2}}+iD\frac{2(\Delta/\gamma)}{\left[  1+(\Delta/\gamma
)^{2}\right]  ^{2}}. \label{eq_D_tilde}%
\end{equation}
The real part of $\mathcal{D}$ corresponds to an actual diffusion of the
polariton. The imaginary part causes quadratic dispersion within the
$k_{\perp}$ spectrum, with a functional form identical to that of the optical
paraxial diffraction, and is thus referred to as motional-induced diffraction (MID).

On resonance $\Delta=0$, the polariton diffusion matches precisely the atomic
diffusion $\mathcal{D}=D$. Besides an overall absorption and phase-shift
originating from $i\chi_{0},$ the evolution of the polariton is a linear sum
of optical diffraction with respect to the distance travelled $\left(
\partial_{z}\Omega\right)  _{\text{diffraction}}=i\nabla_{\perp}^{2}%
\Omega/(2q)$ (due to the polaritons's light constituent) and atomic diffusion
with respect to time $\left(  \partial_{t}\Omega\right)  _{\text{diffusion}%
}=D\nabla_{\perp}^{2}\Omega$ (due to its matter constituent). For the latter,
we translated $v_{g}\partial_{z}\rightarrow\partial_{t}.$ The relative weight
of diffraction and diffusion is thus controlled by the group velocity. Off the
Raman resonance $\Delta\neq0$, the polariton diffusion slows down. The real
part of $\mathcal{D}$ decreases with increasing $|\Delta|$, until vanishing
completely at $\Delta=\pm\gamma.$ At this detuning, the polariton does not
experience any standard diffusion, while the remaining $O(\nabla_{\perp}^{4})$
term gives rise to sub-diffusion evolution.%
\begin{figure}
[tb]
\begin{center}
\includegraphics[
height=4.14cm,
width=8.7784cm
]%
{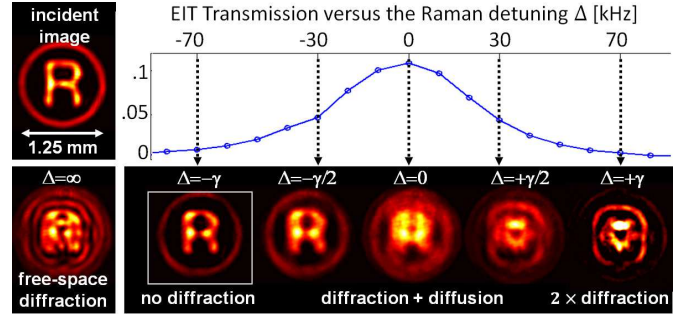}%
\caption{Polariton diffusion and motional-induced diffraction. A Zeeman EIT
setup is used, similar to that in Fig. \ref{fig_diffusion}, with a cell length
$L=5$ cm and optical depth $2\alpha L=6$. The free-space diffraction (bottom
left) is compared to transmitted slow images at several Raman detunings
(right). At $\Delta=0,$ the polariton is delayed by $\sim6$ $\mu$s and
experiences the combination of optical diffraction and diffusion ($D=11$
cm$^{2}$/s for $10$ Torr of neon). At $\Delta<0,$ both diffusion and
diffraction are reduced, and at $\Delta=-\gamma\approx-70$ kHz, they are
completely suppressed ($Dq=v_{g}$ $\protect\cong8700$ m/s). Numerical
calculations confirm that the small difference between the input image and the
transmitted image at $\Delta=-\gamma$ is primarily due to residual $\nabla
^{4}$ terms. At $\Delta=+\gamma,$ no diffusion occurs, but the polariton
experiences the sum of equal optical and motional-induced diffraction, as if
the image has propagated a free-space distance of $2L$. Adapted from
\textcite{FirstenbergNatPhys2009}.}%
\label{fig_diffration}%
\end{center}
\end{figure}

Moreover, at $\Delta\neq0$ the MID becomes nonzero and adds up to the optical
diffraction. The detuning determines the sign of the MID, with
$\operatorname*{Im}(\mathcal{D})>0$ at positive detuning adding to the optical
diffraction, and $\operatorname*{Im}(\mathcal{D})<0$ at negative
detuning\ negating it. While the maximum MID is obtained at $\Delta
=\pm3^{-1/2}\gamma,$ the more interesting case is $\Delta=\pm\gamma$, in which
$\mathcal{D}=\pm iD/2$ is purely imaginary, inducing diffraction without
diffusion. Here the ratio between $\pm D/v_{g}$ and $1/q$ determines the
balance between the optical and induced diffraction, and, for given $D$ and
$q,$ it is governed by the group velocity.%

\textcite{FirstenbergPRL2009}
proposed utilizing MID to completely eliminate the paraxial diffraction in the
medium, by choosing $\Delta=-\gamma$ and $v_{g}=qD$. At these conditions, both
the imaginary and the real parts of the $\nabla_{\perp}^{2}$ coefficient
vanish in Eq. (\ref{eq_effective_diffusion}), rendering a diffraction-less,
diffusion-less, medium. Conversely, at $\Delta=+\gamma$ the actual diffraction
in the medium is twice that in free-space. The non-diffraction condition
$v_{g}=qD$ can intuitively be derived by requiring the diffusion spreading of
a focused Gaussian beam to be equal to its diffraction spreading after one
Rayleigh distance $z_{R}=qw_{0}^{2}/2,$ where $w_{0}$ is the beam
waist-radius. Since the beam does not expand, it is virtually trapped in two
dimensions by the diffusing atoms, in an interesting analogy to the mechanism
of Doppler cooling of atoms by red-detuned light. The latter also relates to a
proposal by
\textcite{ScullyPRL2001}
to stop light propagation using one-photon detuning in a bufferless cell.

\begin{figure}
[tb]
\begin{center}
\includegraphics[
height=1.7411cm,
width=8.7761cm
]%
{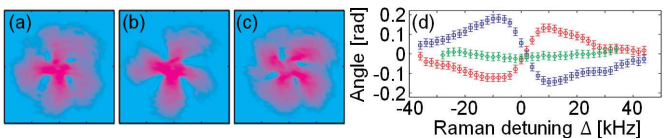}%
\caption{(color online) Collective rotation of a vortex array due to induced
diffraction by \textcite{KatzOE2012}. Hyperfine EIT with $^{87}$Rb is
performed in a cell of length $L=7.5$ cm with $20$ Torr neon ($D=6$
cm$^{2}/s,$ $v_{g}=5000$ m/s). An array of four $m=+1$ vortices (total angular
momentum $J=4$) rotates clockwise at $\Delta=0$ (panel b), the same as it does
in free-space. At (a) $\Delta<0$, the optical diffraction is counteracted,
leading to a counterclockwise rotation with respect to $\Delta=0.$ At (c)
$\Delta>0$, diffraction is enhanced, leading to increased clockwise rotation.
Images are $1$ mm $\times$ $1$ mm. (d) Rotation angle for a two-vortex array
with (red $\bigcirc$) $J=2,$ (blue $\square$) $J=-2,$ and (green $\diamond$)
$J=1-1=0$. All-optical control on the vortices motion could be useful for fast
optical-trapping applications. Adapted from \textcite{KatzOE2012}.}%
\label{fig_katz}%
\end{center}
\end{figure}

These effects were studied by
\textcite{FirstenbergNatPhys2009}
at the condition $v_{g}=qD$ and are all demonstrated in Fig.
\ref{fig_diffration}: The image exhibits optical diffraction (far detuned),
diffusion ($\Delta=0$), non-diffraction ($\Delta=-\gamma$), and double
diffraction ($\Delta=+\gamma$).
\textcite{KatzOE2012}
examined the MID of an array of optical vortices, as shown in Fig.
\ref{fig_katz}.%

As we mentioned earlier, extensive study was devoted to the manipulation of
diffraction by modulating the susceptibility in real space, with either the
coupling beam, the probe beam, or the medium itself inducing the necessary
inhomogeneity of the refraction index\footnote{Electromagnetically-induced
focusing by an inhomogenous coupling field was realized by
\textcite{MoseleyPRL1995}
in hot vapor and by
\textcite{MitsunagaPRA2000}
in a cold ensemble. Schemes in which certain transverse modes evolve without
diffracting due to the non-uniformity of the coupling field were referred to
as induced solitons \cite{WilsonGordonPRA1998}, induced waveguides
\cite{TruscottPRL1999,AgarwalPRA2000}, and transverse confinement
\cite{LukinPRL2005_strong_confinement,chengPRA2005}. Waveguiding was also
demonstrated using the inhomogenous density in a cold atomic cloud
\cite{VengalattorePRL2005,TarhanOL2007}. Lastly, it was proposed that
self-focusing via a Kerr-like effect will support spatial solitons
\cite{HongPRL2003,FriedlerOL2005}.}. In all these schemes, specific transverse
modes are maintained, but a general multi-mode field disperses and may perhaps
regenerate after a certain self-imaging distance \cite{ChengOL2007}. In
contrast, diffraction-manipulation with linear optics in $k_{\perp}-$space, in
the form of Eq. (\ref{eq_effective_diffusion}), applies to multi-mode fields
with arbitrary phase and intensity patterns. Since no actual waveguide is
defined, the medium suspends the expansion of an incoming beam wherever it
impinges on the input plane.

\begin{figure}
[tb]
\begin{center}
\includegraphics[
height=3.6305cm,
width=8.4727cm
]%
{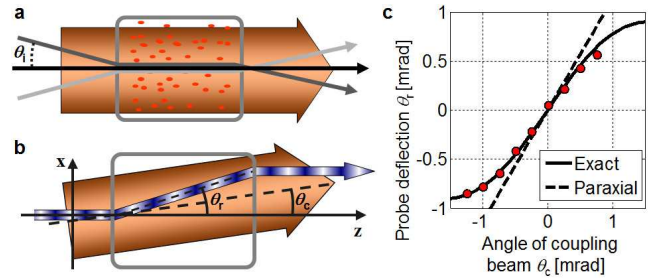}%
\caption{Probe deflection. (a) Non-diffraction can be understood as the
deflection of all rays into a common direction. The rays maintain their phase
relations while traversing the medium and afterwards deflect back into their
original direction. (b) Generally, the probe beam deflects towards the
direction of the coupling beam for $\Delta<0$ and away from the coupling for
$\Delta>0.$ At $\Delta=\pm\gamma,$ a Snell-like law determines the ratio
between $\theta_{r}$ and $\theta_{c}.$ (c) In the non-diffraction conditions,
the probe beam assumes the direction of the coupling beam for small enough
angles ($\circ$ are measurement, line are theory). Adapted from \textcite
{FirstenbergNatPhys2009}.}%
\label{fig_drift}%
\end{center}
\end{figure}

It is instructional to define an \emph{index of diffraction} $n_{\text{diff}%
}=(1-qD/v_{g})^{-1}$, equivalent to the index of refraction as far as paraxial
diffraction is concerned. Without atomic motion ($D=0$), diffraction is not
altered ($n_{\text{diff}}=1$). At the non-diffraction conditions, the index
diverges ($n_{\text{diff}}\rightarrow\infty$). Snell's law, $\sin\theta
_{i}=n_{\text{diff}}\sin\theta_{r}$, then implies no angular divergence inside
the medium $\theta_{r}=0$ regardless of the incident angle $\theta_{i}$ and
hence no diffraction, as illustrated in Fig. \ref{fig_drift}a.%

\begin{figure}
[tb]
\begin{center}
\includegraphics[
height=2.9704cm,
width=7.9654cm
]%
{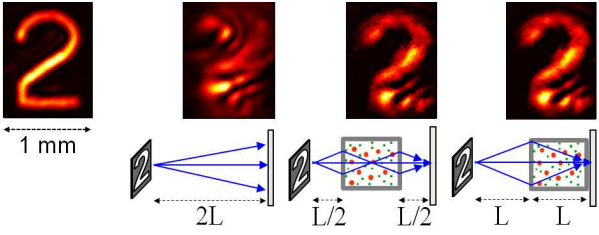}%
\caption{(color online) Manifestation of a negative-diffraction lens with
$n_{\text{diff}}=-1$. The optical diffraction is exactly reversed by setting
$v_{g}=$ $Dq/2$, so that an image diffracting along a distance $L$ before the
cell is re-imaged at the end of the cell. The positive and negative
diffraction 'commute', \textit{e.g.}, it makes no difference if the positive
diffraction occurs partly before and partly after the cell, as exemplified by
the two right images. The restriction $k_{\perp}\ll k_{0}$ was not fulfilled
in this experiment, hence the imperfections in the imaging. Solid arrows
indicate geometrical rays (not $\mathbf{q}$ vectors) refracting oppositely to
the incident angle $\theta_{r}=-\theta_{i}$, according to Snell's law
$\sin\theta_{i}=n_{\text{diff}}\sin\theta_{r}$, as expected in a
negative-index material. Adapted from \textcite{FirstenbergNatPhys2009}.}%
\label{fig_lens}%
\end{center}
\end{figure}

Now, consider the possibility of reducing $v_{g}$ below $qD$, still with
$\Delta=-\gamma,$ so that the (negative) MID be further strengthened. Then,
both the overall diffraction of the polariton and the index of diffraction
become negative. The medium undoes a paraxial diffraction that already took
place, manifesting a negative-index lens in the spirit of
\textcite{Veselago1968}
and
\textcite{PendryPRL2000}%
. Remarkably, the imaging conditions of such a lens are insensitive to its
position between the object and the image, as shown for $n_{\text{diff}}=-1$
in Fig. \ref{fig_lens}.

An important caveat when working at large Raman detunings is the reduced
transmission; even for high coupling intensities ($\gamma=\gamma_{P}$), the
absorption at $\Delta=\pm\gamma$ cannot be rendered lower than
$\operatorname{Im}\chi_{0}\approx\alpha/2.$ This translates to a low
transmission, of about $\exp\left(  -5\right)  ,$ at the Rayleigh distance of
a beam with $w_{0}=\pi/k_{0}$, which is the minimal pixel size allowed under
the $k_{\perp}\ll k_{0}$ condition. The experiments by
\textcite{FirstenbergPRL2009}
took place under these conditions.

\subsection{Induced drift and artificial vector-potential}

The attentive reader may have already realized that, while $n_{\text{diff}}$
alters the refraction at the entrance and the exit of the medium, it is the
direction of the coupling beam that determines the virtual plane of incidence
for this refraction. In fact, since the real index of refraction in dilute
vapor is only marginally different than unity ($n=1\pm10^{-6}$), the actual
entrance plane of the cell has no optical significance. It is thus the virtual
plane perpendicular to the coupling-beam direction which defines the incident
and refraction angles for the modified Snell's law $\sin(\theta_{i}-\theta
_{c})=n_{\text{diff}}\sin(\theta_{r}-\theta_{c}),$ where $\theta_{c}=0$ for an
axial coupling beam. Therefore, tilting the coupling beam results in an
angular deflection of the probe beam in the cell. For a straight-on incidence
($\theta_{i}=0$), the modified Snell's law yields $\theta_{r}=\theta
_{c}(1-n_{\text{diff}}^{-1})$ (see Fig. \ref{fig_drift}b). In particular, at
the non-diffraction conditions ($n_{\text{diff}}\rightarrow\infty$), the probe
deflects exactly onto the direction of the coupling beam ($\theta_{r}%
=\theta_{c}$), as shown in Fig. \ref{fig_drift}c.%

Mathematically, tilting the coupling beam superimposes a transverse phase
grating $\exp(ixq\theta_{c})$ on the Raman interference pattern, replacing
$Dk_{\perp}^{2}$ in Eq. (\ref{eq_k2_k4}) by $D(\mathbf{k}_{\bot}-q\theta
_{c}\mathbf{\hat{x}})^{2}$. For angles small enough ($q\theta_{c}\ll k_{0}$),
the resulting term in Eq. (\ref{eq_effective_diffusion}), $2(\mathcal{D}%
/v_{g})\mathbf{\nabla}_{\bot}\cdot\mathbf{\hat{x}}q\theta_{c},$ induces a
directional deflection on the probe at an angle $\theta_{r}=\mp(qD/v_{g}%
)\theta_{c}=\mp(1-n_{\text{diff}}^{-1})\theta_{c}$, in accordance with the
modified Snell's law. It is worthwhile emphasizing that the deflection effect
does not involve an actual refraction of the optical wavevector ($\mathbf{q}%
$). Similarly to the walk-off phenomenon in birefringent crystals, the wave
fronts (equal phase planes) maintain their original orientation. Hence, the
deflection is unobservable for plane waves and has meaning only for finite
beams. In analogy to a group velocity, which can be modified either via $n$ or
via the dispersion $dn/d\omega$, the deflection here is a (spatial) group
effect, in which the transverse dispersion $\sim dn/dk_{\perp}$ changes the
propagation trajectory. \qquad

In the popular analogy between paraxial light propagation and the
Schr\"{o}dinger dynamics of a massive particle in two-dimensions, the
wavevector plays the role of the mass. When the MID at $\Delta=\pm\gamma$
dominates the optical diffraction, and one translates $z\rightarrow v_{g}t$ in
Eq. (\ref{eq_effective_diffusion}), the effective mass is $m=\pm\hbar/D$. A
phase gradient imposed by the coupling fields thus translates to a vector
potential (VP) for a charged particle:%
\begin{equation}
i\hbar\partial_{t}\Omega(x,y)=\frac{1}{2m}(i\hbar\mathbf{\nabla}_{\perp
}+e\mathbf{A})^{2}\Omega(x,y),
\end{equation}
where $\mathbf{A}=i\hbar\mathbf{\nabla}_{\perp}\ln\Omega_{c}^{\ast}%
(\mathbf{r})$ and $e=1$.

As reviewed by
\textcite{DalibardRMP2010}%
, artificial VP created by the optical dressing of neutral atoms is a major
field of study. Here however, the polaritons, and not the atoms themselves,
experience the artificial VP. As a result, the coupling beam can be used to
mimic the operation of electromagnetic fields on the polariton. In particular,
a tilted coupling beam $\Omega_{c}(\mathbf{r})=\Omega_{c}\exp(ixq\theta_{c})$
produces a uniform VP $\mathbf{A=}\hbar q\theta_{c},$ explaining the
deflection effect via a momentary electric 'kick' at the entrance of the cell
$\mathbf{E}_{\text{in}}=-\partial_{t}\mathbf{A}=\delta(t-t_{\text{in}})\hbar
q\theta_{c}\mathbf{\hat{x},}$ after which the probe propagates in a straight
trajectory. A second kick at the exist face deflects the probe back to its
original direction. Alternatively, a vortex coupling beam with a helical phase
$\exp(im\phi)$ ($\phi$ the azimuthal angle) inflicts a kick in the azimuthal
direction. The underlying VP $\mathbf{A=}\hbar m\mathbf{\nabla}_{\perp}\phi$
implies an artificial magnetic field $B=\partial_{x}A_{y}-\partial_{y}%
A_{x}=2\pi\delta(x)\delta(y)\hbar m$ along the dark vortex core, whereas the
probe can only propagate at the brightened areas around the core. Altogether,
a probe in the form of a ring of lobes is predicted to rotate while
propagating in the medium and cease rotating upon exiting
\cite{YankelevThesis2012}.

\section{Coherent diffusion of stored light\label{sec_storage}}

\begin{figure}
[tb]
\begin{center}
\includegraphics[
height=3.6944cm,
width=8.1653cm
]%
{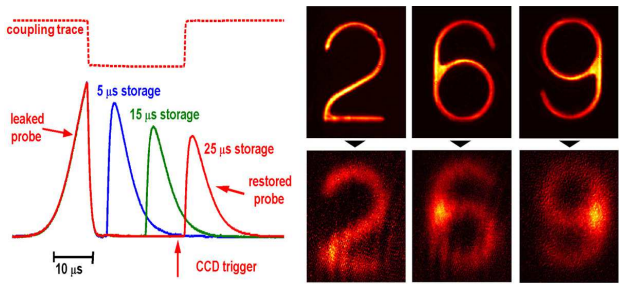}%
\caption{(color online) Diffusion during storage-of-light in vapor. Left: The
first half of the probe pulse is allowed to leak before storage. The second
half is stored by turning off the coupling beam for a duration $\tau$, after
which the probe revives. The traces show the probe for $\tau=5,$ $15$ and $25$
$\mu$s, and the coupling for $\tau=25$ $\mu$s. Right: The spatial effect of
diffusion is observed by comparing the images of the input probe (top row) and
the retrieved probe (bottom row). Storage durations are $\tau=2$, $6$, and $9$
$\mu$s. Adapted from \textcite{ShukerImaging2007}.}%
\label{fig_digits}%
\end{center}
\end{figure}

Diffusion and diffraction of dark-state polaritons, discussed in the previous
section, arise from the interplay between the atomic motion and the
propagating excitation. Perhaps more elementary is the effect of the atomic
motion on the atomic coherence in the absence of the light, as occurs during
light storage. In light storage, the polariton is transformed into a
matter-only excitation which does not propagate. The ground-state atomic
coherence stores the light amplitude in the form of a spatial spin-wave, later
to be mapped back to a propagating polariton. Storage of light is accomplished
with EIT by switching-off the coupling beam --- and switching it back on for
retrieval \cite{HauNature2001,PhillipsPRL2001}, see Figure \ref{fig_digits}
(left). Storage and retrieval can also be performed with a longitudinal
gradient of the frequency detunings. This method, known as gradient-echo
memory (GEM), was recently implemented with ground-state coherence in a
$\Lambda-$system \cite{BuchlerOL2008}.

When light storage is performed with a single quantum, ideally by storing a
single photon, it realizes a quantum memory --- a fundamental building block
for quantum communication and computation \cite{DuanNature2001,PolzikRMP2010}.
In atomic ensembles, the single quantum is stored in the collective state of
all atoms. Unconditional storage of light on the level of single-photons was
recently achieved by
\textcite{BuchlerNatPhys2011}
using GEM in a hot buffered cell. However, most of the experiments so far have
used spontaneous Raman scattering to generate the spin wave, conditioned on
the detection of a scattered photon
\cite{ChouPRL2004,EisamanNature2005,MatsukevichPRL2006,KuzmichNatPhys2008,BashkanskyOL2012}%
. Diffusion of the atoms before the spin wave is converted back to light poses
the same issues as in unconditional storage, as we describe in this section.%

\subsection{Diffusion of a stored coherence field}

When storage is performed, the three-dimensional spatial envelope of the probe
$\Omega(\mathbf{r})$ is linearly mapped onto the ground-state
coherence\footnote{Alternatively, in spontaneous storage, the superposition of
the coupling beam and a spontaneously generated photon heralding the storage
is saved on the coherence field $\rho_{12}(\mathbf{r},\tau=0)$.} $\rho
_{12}(\mathbf{r},\tau=0)$. The dynamics during the storage time $\tau$ is
governed by%
\begin{equation}
\partial_{\tau}\rho_{12}(\mathbf{r},\tau)=D(\mathbf{\nabla}+i\mathbf{k}%
)^{2}\rho_{12}(\mathbf{r},\tau)-\gamma_{0}\rho_{12}(\mathbf{r},\tau),
\label{eq_diffusion}%
\end{equation}
which derives from Eq. (\ref{eq_rho21}) in the absence of light. Even for a
uniform envelope and negligible damping \textbf{(}$\mathbf{\nabla}%
\rightarrow0,$ $\gamma_{0}\rightarrow0$), the diffusion of atoms through the
Raman wave results in a dephasing of rate $Dk^{2}.$
\textcite{FelischhauerPRA2002}
and
\textcite{MewesFleischhauer2005}
show that the decoherence of the quantum memory (in terms of the fidelity of
the stored state) is proportional to this dephasing.\qquad

In a recent experiment,
\textcite{PanNatPhys2008}
showed that the memory time in a cold atomic gas reduces with the angle
between the Raman beams and is determined by the time it takes the atoms to
(ballistically) move one Raman wavelength ($\tau_{d}\varpropto k^{-1}%
\propto\theta^{-1}$). Indeed, memory times as long as milliseconds were
achieved by
\textcite{PanNatPhys2008}
and by
\textcite{HauPRL2009}
using collinear beams ($k\approx0$), and by
\textcite{KuzmichNatPhys2008}
with an optical trap that confines the atomic motion in the direction of
$\mathbf{k}$. Furthermore,
\textcite{BlochPugatchPRL2009}
demonstrated light storage with ultra-cold atoms trapped in a
three-dimensional optical lattice (a Mott insulator). The confinement of
atomic motion to a site much smaller than the optical wavelength allowed
Schnorrberger \textit{et al.} to imprint phase gradients of wavenumbers
$k\mathbf{=}$ $\theta q,$ with $\theta$ as large as 25 mrad, while maintaining
the memory for more than $0.1$ ms. All this of course does not apply to a BEC
where, due to its long-range coherence, stored light was retrieved even after
the atoms moved numerous $\lambda_{R}$ \cite{HauNature2007}.

Nevertheless, even when $k\approx0$, atomic motion plays an important role in
the storage of finite-size and structured fields. Diffusion of the atomic
coherence\footnote{Note that the ground-state populations, $\rho
_{11}(\mathbf{r})$ and $\rho_{22}(\mathbf{r})$, diffuse in a similar manner,
but, in the weak-probe regime, their contribution to the storage is small.}
$\partial_{\tau}\rho_{12}=D\nabla^{2}\rho_{12}$ can be observed directly by
comparing the input image to the retrieved image at different storage
durations. This is especially true when the propagation time before and after
storage is much shorter than the storage duration itself, as is often the
case. Figure \ref{fig_digits} (right) presents measurements of diffusion with
stored images \cite{ShukerImaging2007}. Diffusion is clearly observed by the
smearing of the digits' image and is more pronounced as the storage duration
increases. The spreading of stored information was used by
\textcite{ZibrovPRL2002}
to perform storage and retrieval at two distant locations in the cell. As a
complementary concept,
\textcite{WalsworthJMO2005}
demonstrated two retrievals from the same location due to diffusion of
coherence out and back into the beam area.%
\begin{figure}
[tb]
\begin{center}
\includegraphics[
height=3.4732cm,
width=8.4727cm
]%
{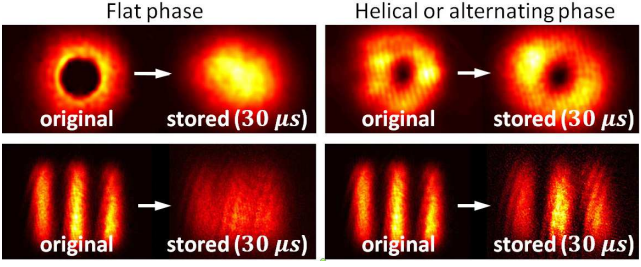}%
\caption{(color online) Diffusion of coherence fields with uniform and
nonuniform phase patterns. A flat-phased ring (top left) is filled-up after a
short storage duration, while a vortex ring with a helical phase is preserved
(top right). Similarly, the blurring of a line pattern (bottom left) can be
reduced by flipping the phase between adjacent lines (bottom right). As
evident, the visibility of the lines with the alternating phase remains much
higher than the flat-phased image. The vortex radius is $670$ $\mu$m, the
lines are $1.5/$mm, and $D=10$ cm$^{2}/$s. Adapted from \textcite
{PugatchPRL2007,ShukerImaging2007}.}%
\label{fig_phase_storage}%
\end{center}
\end{figure}

Let us now take an ideal case with $k=0$ and a coupling beam that covers the
whole medium. Naively it might seem that the total power of the restored probe
is not altered by diffusion, as diffusion is a conserving process, $\int
\rho_{12}=$\textit{const}. However, it is the light-field amplitude
$\Omega\propto\rho_{12}$, rather than its intensity $|\Omega|^{2}$, that
effectively diffuses, and the total power $P\propto\int|\Omega|^{2}$ decays.
For example, a stored Gaussian beam that doubles its area due to diffusion
conveys a half of its initial power. This geometric effect was shown to limit
the storage time of images and of narrow beams in buffered cells
\cite{ShukerImaging2007, BuchlerNatComm2011,LettOE2012}.

In contrast to standard 'heat' diffusion, stored 'images' can be
complex-valued, as the phase pattern of the probe is exactly imprinted on the
diffusing coherence \cite{FleischhauerPRL2000}. Patterned phase leads to
effects of constructive and destructive interference during diffusion, similar
to those occurring in light propagation. For instance, consider the diffusion
of the annular ring shown in Fig. \ref{fig_phase_storage} (top). A flat-phased
ring is completely filled up after a short storage time, while the dark center
of a stored vortex (LG$_{01}$ mode) is well maintained. The vortex core
remains dark due to destructive interference: the phase around the dark center
completes a 2$\pi$ twist, and the contributions of all atoms diffusing inwards
sum up to zero. Similar behavior is achieved by applying a well-designed phase
pattern on specific images. The blurring of three resolution lines in Fig.
\ref{fig_phase_storage} (bottom) is reduced by flipping the phase between
adjacent lines. The decay of the lines' visibility due to diffusion is
dramatically slowed down. The same principles are used in optical phase-shift
lithography to overcome the diffraction limit of small adjacent features. The
downside of using destructive interference is the faster decay of the total
retrieved power. The decay rate increases with the complexity of the phase
pattern, thereby decreasing the fidelity of the retrieved states
\cite{YelinPRA2008b}.

\begin{figure}
[tb]
\begin{center}
\includegraphics[
height=3.6349cm,
width=7.3585cm
]%
{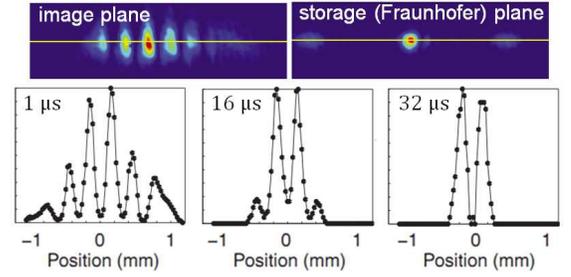}%
\caption{(color online) By storing the Fraunhofer diffraction pattern (top
right) instead of the image itself (top left) the fine details of the image
are better preserved under diffusion (bottom). Rather, diffusion attenuates
the outer parts of the beam. Adapted from \textcite{HowellPRL2008}.}%
\label{fig_howell}%
\end{center}
\end{figure}

An alternative method utilizing the phase pattern for the suppression of
diffusion was analyzed by
\textcite{YelinPRA2008}
and realized by
\textcite{HowellPRL2008}
(Fig. \ref{fig_howell}). Instead of the image itself, Zhao \textit{et al.}
suggested storing the Fraunhofer diffraction pattern at the center of a $4f$
telescope. Rather than filtering the high spatial components (a convolution
with a Gaussian), diffusion merely attenuates the outermost parts of the
original image (Gaussian multiplication), thereby maintaining its fine
details. In contrast with the phase-shift method, no \textit{a-priori}
information about the image is required.

\subsection{Shape-preserving modes of coherent diffusion\label{ss_modes}}

In free-space optics, the paraxial-diffraction equation $\partial_{z}%
\Omega=i\nabla_{\perp}^{2}\Omega/(2q)$ has several sets of shape-preserving
solutions. These are notably the polynomial-Gaussian modes, including the
well-known \emph{standard} Hermite-Gauss (sHG) or Laguerre-Gauss (sLG) modes.
Their transverse intensity pattern is maintained along the propagation
direction $z$ and scaled according to the beam radius $w_{z}=w_{0}%
\sqrt{1+(z/z_{R})^{2}},$ where $z_{R}=qw_{0}^{2}/2$ is the Rayleigh distance.
For example, the sHG mode $E_{n,m}^{\text{sHG}}(x,y,z;w_{0})$ has the form%
\begin{equation}
H_{n}\left(  \sqrt{2}\frac{x}{w_{z}}\right)  H_{m}\left(  \sqrt{2}\frac
{y}{w_{z}}\right)  \exp\left(  -\frac{x^{2}+y^{2}}{\tilde{w}_{z}^{2}}\right)
,\nonumber
\end{equation}
where $\tilde{w}_{z}=\sqrt{2(z_{R}-iz)/q}$ is the complex radius and $H_{k}$
the Hermite polynomials. $N=n+m$ is the total mode order.%
\begin{figure}
[tb]
\begin{center}
\includegraphics[
height=6.3816cm,
width=7.9654cm
]%
{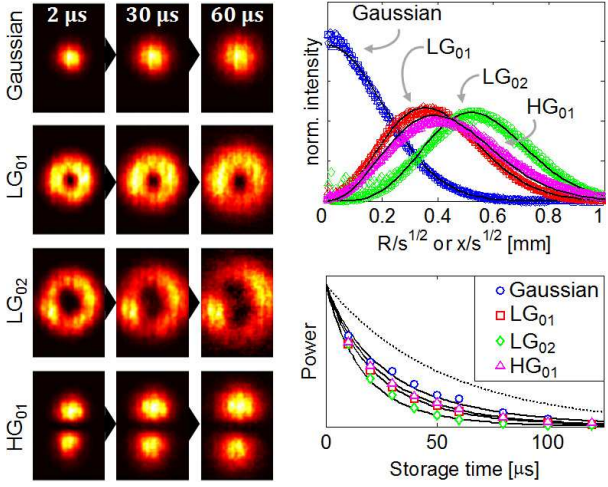}%
\caption{(color online) Diffusion of 'common' modes during storage. Left:
Intensity patterns of (top to bottom) Gaussian, LG$_{01}$, LG$_{02}$, and
HG$_{01}$, expanding due to diffusion while preserving their shape. Top right:
Normalized cross-sections at different storage durations, $\tau=$($\bigcirc$)
2, ($\square$) 20 , ($\triangle$) 40, ($\diamond$) 60 $\mu s,$ rescaled
horizontally by the stretching factor $s(\tau)$. Bottom right: Decay of the
total power, increasing with the total mode order $N$. Dashed line is
$e^{-2\gamma\tau},$ solid lines are $e^{-2\gamma\tau}s(\tau)^{-2(N+1)}.$
Adapted from \textcite{FirstenbergPRL2010}.}%
\label{fig_modes}%
\end{center}
\end{figure}

A less familiar solution for paraxial diffraction is the set of \emph{elegant}
modes, first studied by
\textcite{SiegmanBook}
for their neater mathematical form. The elegant Hermite-Gauss (eHG) mode
$E_{n,m}^{\text{eHG}}(x,y,z;w_{0})$ has the form%
\[
H_{n}\left(  \frac{x}{\tilde{w}_{z}}\right)  H_{m}\left(  \frac{y}{\tilde
{w}_{z}}\right)  \exp\left(  -\frac{x^{2}+y^{2}}{\tilde{w}_{z}^{2}}\right)  .
\]
Contrast this with the standard mode above, here the polynomial and the
Gaussian have a mutual (complex) scaling, and the $\sqrt{2}$ in the polynomial
argument is absent. A corresponding elegant form for the circular-symmetric LG
modes also exists.

The elegant modes are not shape-preserving in free-space optics and are thus
rarely used. Remarkably, at the focal plane ($z=0$), they were found to be the
basis for the shape-preserving solutions of coherent diffusion in two
dimensions \cite{FirstenbergPRL2010}. Substituting $E_{n,m}^{\text{eHG}%
}\Rightarrow\rho_{12}$ in Eq. (\ref{eq_diffusion}) with $\mathbf{k}=0$, one
finds%
\begin{equation}
E_{n,m}^{\text{retrieved}}(\tau)=e^{-\gamma_{0}\tau}s\left(  \tau\right)
^{-(N+1)}E_{n,m}^{\text{eHG}}(x,y,z=0;w_{\tau}) \label{eq_intensity_decay}%
\end{equation}
where $w_{\tau}=w_{0}s\left(  \tau\right)  $ is the expanding waist radius and
$s\left(  \tau\right)  =(1+4D\tau/w_{0}^{2})^{1/2}$ is the stretching factor.
The shape is therefore preserved, while expanding, throughout the diffusion.
The algebraic decay $s\left(  \tau\right)  ^{-2(N+1)}$ of the total power
$P\propto\int|E|^{2}$, explicated previously for the Gaussian ($N=0$) case,
becomes faster with increasing mode order due to interference between atoms
diffusing through the oscillating phase patterns.

Note that the standard and elegant sets differ only in their polynomial terms,
and therefore low-order HG and all vortices (LG$_{p=0}$) are common to both
sets and preserve their shape under the simultaneous action of diffusion and
diffraction. They are thus the natural modes for slow light --- a result which
is standard and elegant, in both meanings of the words.

\begin{figure}
[tb]
\begin{center}
\includegraphics[
height=3.4024cm,
width=5.8921cm
]%
{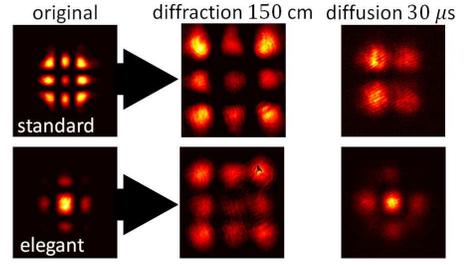}%
\caption{(color online) Evolution of (top row) sHG$_{22}$ and (bottom row)
eHG$_{22}$, measured by \textcite{Yankelev2012}, for (center column)
diffraction in free space and (right column) diffusion during storage of
light. The standard mode is only preserved during diffraction, whereas the
elegant is only preserved during diffusion. Adapted from \textcite
{Yankelev2012}.}%
\label{fig_elegant}%
\end{center}
\end{figure}

The diffusion of low-order (common) LG and HG modes during light storage in
EIT is presented in Fig. \ref{fig_modes}. After scaling and normalization, the
cross-sections at different storage durations of each of the modes collapse to
a single curve.
\textcite{BuchlerArxiv2012}
used GEM to demonstrate the shape-preserving evolution of HG modes and the
associated algebraic decay. By probing an optically pumped medium, the
diffusion of LG beams with radial or azimuthal polarization (vector beams) was
observed by
\textcite{FatemiOL2011}%
.\textsl{ }%
\textcite{Yankelev2012}
experimented with the high-order modes sHG$_{22}$ and eHG$_{22}$ (Fig.
\ref{fig_elegant}). As expected, the shape of the sHG$_{22}$ mode is preserved
during diffraction while dramatically altering during diffusion, and
vice-versa for the eHG$_{22}.$%

\begin{figure}
[tb]
\begin{center}
\includegraphics[
height=3.5627cm,
width=8.1653cm
]%
{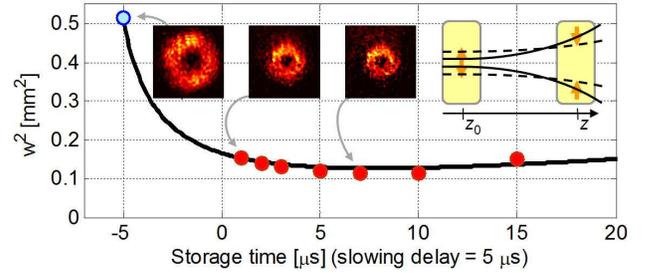}%
\caption{(color online) Shape-preserving contraction and subsequent expansion
of LG$_{01}$ during storage. Due to destructive interference, the diffusion of
a diverging beam initially decreases the beam area $w_{\tau}^{2}$ (line is the
theoretical prediction). Adapted from \textcite{FirstenbergPRL2010}.}%
\label{fig_shrink}%
\end{center}
\end{figure}

Quite an interesting effect occurs when diffusion is performed away from the
focal plane of the beam. The radial phase-oscillations in the transverse
plane, due to the curved phase-fronts of the diverging beam, lead to
destructive interference at the outskirts of the beam during diffusion. The
result is a shape-preserving contraction of the beam, as shown in Fig.
\ref{fig_shrink}, in contrast to the obvious consequence of diffusion. In
effect, diffusion acts to (virtually) expand the waist radius at the focal
plane ($z=0$), even if this plane lies outside the medium, which leads
initially to contraction at $|z|>z_{R}$ (see sketch in Fig. \ref{fig_shrink}).
This effect is directly related to the contraction of slow light out of focus,
presented in Fig. \ref{fig_dicke_amplitude}.%

\section{Finite-size beams, Ramsey narrowing\label{sec_Ramsey}}

Up until now, we have mainly considered a large and uniform coupling beam,
such that any inhomogeneity experienced by the atoms was set by the weak
probe. In fact, the atoms are constantly driven towards the dark state by the
perpetual coupling field, and those that are slow enough can adiabatically
follow the local dark-state $\propto\Omega_{c}^{\ast}\left\vert 1\right\rangle
-\Omega^{\ast}(\mathbf{r})\left\vert 2\right\rangle .$ However this situation
is not prevalent, especially when the Raman fields (in a single or two beams)
have a more symmetric role, such as in CPT and NMOR. There is often a finite
'bright' region, covered by the light, and a remaining large 'dark' region.
The atomic motion within these regions and between them is the subject of this section.

Finite excitation times of ground-state coherence is a well-studied phenomena,
as described by
\textcite{Gawlik1986}
and
\textcite{ArimondoPRA1996}%
, with the atoms either spatially leaving the illuminated area or shifting out
of resonance due to some inhomogenous mechanism. The observed spectra are more
elaborate than those we have studied hitherto, because the finite pumping time
rules out the linear response assumption. Instead of an instantaneous pumping
action, the process becomes kinetic, with different atomic trajectories
contributing differently to the spectra. An example of a non-Lorentzian,
cusp-like spectrum, was presented by
\textcite{Pfleghaar1993}%
. Pfleghaar \textit{et al.} fully described the spectrum by using an exact
geometrical transit-time model, taking into account the possible atomic
trajectories through the inhomogenous beam. Trajectories with a transit time
short compared to the pumping and damping rates $\tau_{t}\ll\gamma_{0}%
^{-1},\gamma_{P}^{-1}$ contribute to the transit-time-limited broad feature;
trajectories with long transit time contribute to the narrower central part of
ultimate width $\gamma\approx\gamma_{0}+\gamma_{P}.$ We note here that
non-Lorentzian spectra also arise for atoms at rest, when non-uniform power
broadening dominates \cite{HollbergPRA2004} .%
\begin{figure}
[ptb]
\begin{center}
\includegraphics[
height=3.4976cm,
width=8.7784cm
]%
{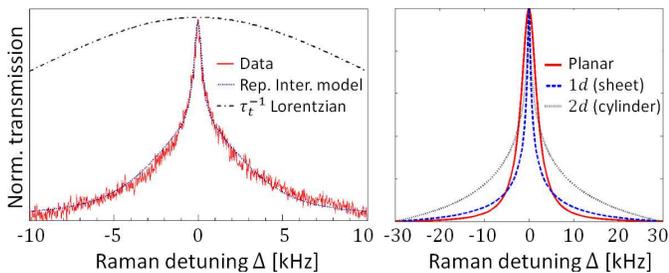}%
\caption{(color online) Ramsey narrowing. Left: Dark-resonance spectrum for
rubidium in 5 Torr neon with 1-mm-wide beam, measured by \textcite
{WalsworthPRL2006} and calculated using the repeated-interactions model by
\textcite{XiaoOE2008}. The comparison to the transit-time ($\tau_{t}^{-1}$)
broadening highlights the substantial Ramsey narrowing. Right: The diffusion
solutions (\ref{eq_ramsey_1d2d}) by \textcite{FirstenbergPRA2008}, for
$2a=0.2$ mm, $D=10$ cm$^{2}$/s, no walls, and large power-broadening
$\gamma_{p}=20\gamma_{0}=2$ kHz. Ramsey narrowing produces a central feature
narrower than $2\gamma_{p}.$ Adapted from \textcite
{XiaoOE2008,FirstenbergPRA2008}.}%
\label{fig_ramsey_narrowing}%
\end{center}
\end{figure}

Coherently pumped atoms that have left the beam may return in a later time
before losing their coherence. Coherent recurrence occurs in wall-coated or
buffered cells, and has long been known as a narrowing mechanisms in standard
rf spectroscopy \cite{Robinson1982_ultra_narrow}. While the homogenous damping
rate ($\gamma_{0}$) sets a lower limit on the width of any spectral feature,
transit-time broadening ($\tau_{t}^{-1}$) is reduced by recurring atoms that
effectively increase the interaction time, and power broadening ($\gamma_{P}$)
is reduced because the recurring atoms have evolved predominantly in the dark.
The initial pumping of the atoms in the bright region, the subsequent
evolution in the dark, and their contribution to the spectrum upon return,
correspond to the Ramsey method of separated oscillating fields
\cite{Ramsey1950}. The associated narrowing was therefore named \emph{Ramsey
narrowing}. For all-optical Raman resonance, Ramsey narrowing was first
observed in wall-coated cells with NMOR \cite{KanorskyAPB1995,BudkerPRL1998}
and EIT \cite{KleinJMO2006}. In both processes, the spectrum exhibits a broad
pedestal feature, attributed to single-transit trajectories, and a narrow
peak, due to coherent atoms returning after long times in the dark
\cite{BudkerRMP2002,budker2005}. Diffusion-induced Ramsey narrowing in
buffered cell was observed in various Raman processes
\cite{ZibrovOL2001,AlipievaOptLett03,NovikovaJOSAB2005,WalsworthJMO2005}, as
exemplified in Fig. \ref{fig_ramsey_narrowing} (left).

The difficulty of writing a linear susceptibility in the form of Eq.
(\ref{chi_bar_dicke}) originates from the nonlinear terms in Eq.
(\ref{eq_rho21}). Even for negligible power-broadening $\gamma_{P}%
(\mathbf{r})\rightarrow0$, the source term $\Omega_{c}^{\ast}(\mathbf{r}%
)\Omega(\mathbf{r})$ yields a convolution in $k-$space that, although
accurate, makes it hard to solve for the spectrum. The following two
approaches to calculate the spectrum thus stay in real space.%
\begin{figure}
[ptb]
\begin{center}
\includegraphics[
height=3.3758cm,
width=8.7784cm
]%
{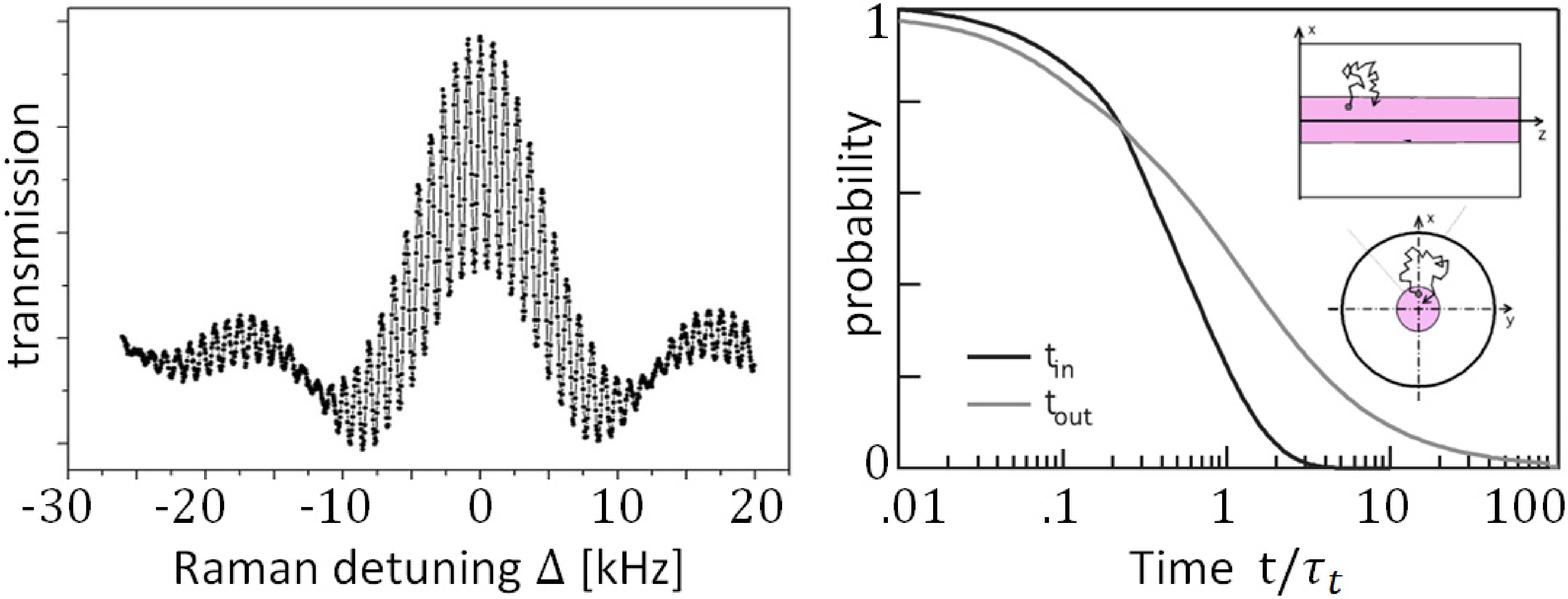}%
\caption{(color online) Left: Ramsey fringes in a dark-resonance, obtained by
\textcite{ClaironPRL2005} with $80$-$\mu$s pulses separated by $1$ ms. Right:
Probability distribution of the durations in the beam ($t_{\text{in}}$) and in
the dark ($t_{\text{out}}$), calculated by \textcite{WalsworthPRL2006} for
atoms diffusing in a cylindrical geometry. Adapted from \textcite
{ClaironPRL2005,WalsworthPRL2006}.}%
\label{fig_fringes}%
\end{center}
\end{figure}

\subsection{Repeated interaction}

Following the original ideas by
\textcite{FrueholzJPB1985}%
, the repeated-interaction model builds the spectrum from an ensemble average
of stochastic atomic trajectories, as delineated by
\textcite{XiaoOE2008}%
. Trajectories may comprise a single transit time ($t_{in}$), a Ramsey process
($t_{\text{in}}/t_{\text{out}}/t_{\text{in}}$), or any longer sequence
($t_{\text{in}}/t_{\text{out}}/t_{\text{in}}/t_{\text{out}}/t_{\text{in}%
}\cdots$). The contribution of longer trajectories is smaller due to the
constant damping $\gamma_{0},$ and the sum thus converges. During the dark
period, the ground-state dipole oscillates at the Raman-detuning frequency
$\Delta$ with respect to the beating frequency of the Raman beams. An atom
leaving the beams in a perfect dark-state will have the probability
$\operatorname{Re}[e^{-(i\Delta+\gamma_{0})t_{\text{out}}}]$ of returning
in-phase (in the dark state), resulting in Ramsey fringes with respect to
$\Delta$. These can be measured by a fixed pulse sequence as shown in Fig.
\ref{fig_fringes} (left). As with Ramsey spectroscopy, the fringes' period is
set by the dark time $t_{\text{out}}^{-1}$ and their envelope by the bright
time $t_{\text{in}}^{-1}.$ Ramsey fringes were observed with Raman processes
by separations in the velocity, time, and space domains
\cite{Buhr1986,SchuhOptComm1993,ZibrovPRA2001,ClaironPRL2005}.

Due to the distribution of the times spent in the bright and dark areas, a
weighted average of such Ramsey fringes constitutes the spectrum.
\textcite{WalsworthPRL2006}
calculated the time probability-distribution of staying in the bright area
$P_{\text{in}}(t)$ and dark area $P_{\text{out}}(t)$ for atoms diffusing
through a cylindrical beam (Fig. \ref{fig_fringes}, right). Similar analysis
for ballistic motion in wall-coated cells was carried out by
\textcite{KleinPRA2011}%
. The calculations assume two spatial dimensions, as the process is virtually
insensitive to the axial motion of the atoms. If the dipole amplitude $A$ of
an atom leaving the beam was fixed, the ensemble average would have read%
\begin{figure}
[ptb]
\begin{center}
\includegraphics[
height=3.4467cm,
width=8.7784cm
]%
{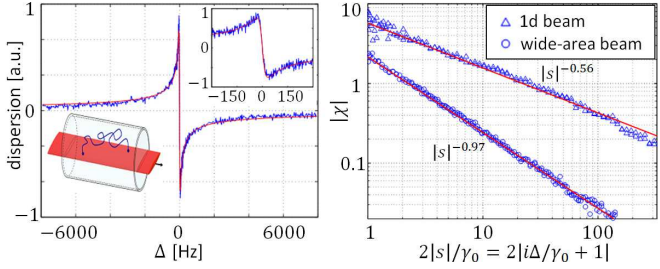}%
\caption{Dispersion spectrum with a one-dimensional light-sheet of width $126$
$\mu$m. Right: Universal power-lows with exponents $-1$ (Lorentzian) and
$-0.5$ (one-dimensional recurrence). Adapted from \textcite{PugatchPRL2009}.}%
\label{fig_universal}%
\end{center}
\end{figure}
\begin{equation}
A\int_{0}^{\infty}dtP_{\text{out}}(t)e^{-(i\Delta+\gamma_{0})t}=AP_{\text{out}%
}(s), \label{eq_Pout_laplace}%
\end{equation}
where $s=i\Delta+\gamma_{0}$ and $P_{\text{out}}(s)$ is Laplace transform of
$P_{\text{out}}(t).$ The full repeated interaction model involves nested
integrals essentially similar to that of Eq. (\ref{eq_Pout_laplace}). To
calculate the more intricate evolution in the bright stages, which involves
dark-state pumping,
\textcite{XiaoOE2008}
use the 3-element vector model by
\textcite{ShahriarPRA1997}%
. The model reduces the master equation of the density matrix into a set of
three Bloch equations, under the assumption of negligible $\gamma_{0}$. The
evolution of the reduced vector has a closed mathematical solution in the form
of a damped precession. A Ramsey sequence is then obtained by chaining three
(in/out/in) solutions.
\textcite{XiaoOE2008}
generalize the vector model to account for finite $\gamma_{0}$ and obtain an
analytic expression for all Ramsey spectra. Integrating over the trajectories
using $P_{\text{in}}(t)$ and $P_{\text{out}}(t)$, the reconstructed spectrum
agrees very well with the measurements (Fig. \ref{fig_ramsey_narrowing}, left)
for a range of experimental parameters.
\textcite{KleinPRA2011}
augment the model with a forth atomic state, to account for optical pumping
out of the $\Lambda-$system due to strong light fields. Indeed, for both
ballistic and diffusing atoms, the distribution of bright times turns the
Ramsey envelope into a broad spectral feature, while the distribution of dark
times wipes out the Ramsey fringes, leaving a single pronounced narrow feature
at the line center.

Recently,
\textcite{PugatchPRL2009}
analyzed the limit of an infinitely small beam, for which the transit-time
broadening, and hence the fringes envelope, is very large. Since
$P_{\text{in}}(t>0)\rightarrow0,$ the bright periods have a negligible effect
on the spectrum, which becomes independent of the beam size and, in that
respect, universal. While the atoms are essentially always in the dark, a
non-zero ground-state dipole is sustained by the weak beam. The average dipole
is given by an infinite sum of multiple periods in the dark, each one given by
Eq. (\ref{eq_Pout_laplace}), $A\sum_{n}P_{\text{out}}(s)^{n}%
=A/[1-P_{\text{out}}(s)].$ As evidenced by this expression, the resulting
complex spectrum, measured by
\textcite{PugatchPRL2009}
(Fig. \ref{fig_universal}, left), constitutes a direct signature of the time
distribution in the dark. Moreover, as the beam is infinite small,
$P_{\text{out}}(t)$ is equivalent to the so-called first return-time
distribution FRT$(t),$ which is the universal probability distribution for a
random walker of returning to the origin at time $t.$ In one dimensional
diffusion, corresponding to the sheet-like beam used in the experiment,
FRT$(s)=P_{\text{out}}(s)=(4Ds)^{-1/2},$ in striking contrast to the complex
Lorentzian spectrum $\varpropto s^{-1}.$ These power-low decays are shown in
Fig. \ref{fig_universal}, right.

\subsection{Diffusion solution}

Although providing insight into the Ramsey-narrowing process, the
repeated-interaction model applies the same physics already contained in the
diffusion-equation formalism of the previous sections. One can essentially
obtain the spectra from the coupled internal and external dynamics of the
density-matrix distribution. To this end, we express the spatially dependent
source and pumping rates, $S(\mathbf{r},t)\ $and $\gamma_{P}(\mathbf{r})$ in
Eq. (\ref{eq_rho21}), using the profiles of the beams, $\Omega(x,y)$ and
$\Omega_{c}(x,y)$, and then solve the diffusion equation for the steady-state
distribution of the ground-state dipoles $\rho_{21}(x,y).$ The optical dipole
$\rho_{31}(x,y)$ is calculated from Eq. (\ref{eq_rho31}), and an integration
over the beam profile yields the absorption spectrum $P\varpropto
\operatorname{Im}\int dxdy\Omega^{\ast}\rho_{31}.$ As a matter of fact, such
mathematical procedure conflicts with a previous notion, that steady-state
solutions cannot accurately describe transit-time-limited spectra
\cite{Gawlik1986}.%

\textcite{XiaoOE2008}
wrote a similar diffusion equation using the 3-element vector model and by
that generalized Eq. (\ref{eq_rho21}) to include a non-weak probe --- and
essentially any ratio between the Raman beams, including the balanced case.
Numerical solution of the diffusion equation in this model, for a small
Gaussian beam, was shown by Xiao \textit{et al.} to agree with the
repeated-interaction model.

For a few simple geometries, it is possible to obtain closed-form expressions
for the spectra, as corrections $R\left(  \Delta\right)  $ to the stationary
spectrum $\chi_{0}\rightarrow\chi_{0}(1-R)$. For a stepwise light sheet (1d)
or a top-hat beam (2d) of widths $2a$, and absorbing boundary conditions at
the walls at a distance $b,$ the diffusion solution gives
\begin{align}
R^{\text{1d}}\left(  \Delta\right)   &  =\frac{1}{\kappa a}\frac{\tanh(\kappa
a)}{1+\left(  \kappa/\kappa_{0}\right)  \tanh\left(  \kappa a\right)
\tanh[\kappa_{0}\left(  b-a\right)  ]},\nonumber\\
R^{\text{2d}}\left(  \Delta\right)   &  =\frac{2}{\kappa a}\left[  \frac
{I_{0}\left(  \kappa a\right)  }{I_{1}\left(  \kappa a\right)  }+\frac{\kappa
}{\kappa_{0}}\frac{K_{0}\left(  \kappa_{0}a\right)  }{K_{1}\left(  \kappa
_{0}a\right)  }\left(  1-\beta\right)  \right]  ^{-1}, \label{eq_ramsey_1d2d}%
\end{align}
where $\beta=K_{0}\left(  \kappa_{0}b\right)  K_{0}^{-1}\left(  \kappa
_{0}a\right)  I_{0}^{-1}\left[  \kappa_{0}(b-a)\right]  $ is due to the walls.
Here, $\kappa$ and $\kappa_{0}$ are defined via $D\kappa^{2}=\gamma_{0}%
+\gamma_{p}-i\Delta$ (inside the beam) and $D\kappa_{0}^{2}=\gamma_{0}%
-i\Delta$ (outside), and $I_{n}$, $K_{n}$ are the modified Bessel functions.
These expressions revert to the transit-time limit for a circumferential wall
($b=a$) that depolarizes all atoms before they recur. The solution with no
walls ($b\rightarrow\infty$) was presented by
\textcite{FirstenbergPRA2008}
and shown in Fig. \ref{fig_ramsey_narrowing} (right); The reduction of power
broadening is clearly visible on the central feature. One may also recovers
the asymptotic universal behavior shown in Fig. \ref{fig_universal} by taking
$a\rightarrow0.$ Finally, minor corrections for non-flat beams where solved
by
\textcite{YatsenkoQuantPh2008}%
.

\section{Outlook}

We have presented the physics of Raman processes with hot atoms, whose
internal coherence is preserved despite their external motion. The unique
combination of rapid atomic motion, large Raman wavelengths, long lifetimes,
and large group delays, was shown to have diverse, significant spectral and
spatial consequences. The same physical principles hold for a rich variety of
Raman schemes and matter systems that are either out of the scope of this
Colloquium or yet to be explored.

The spectra we have been studying derive from the exponential or Gaussian
dephasing rate, pertaining to regular thermal motion. In two dimensional
systems, power-law decay of the velocity correlation is manifested by
L\'{e}vy-like Raman spectra, whereas more intriguing spectra are expected for
non-equilibrium one dimensional systems. These are realizable with cold atoms,
for which it is also exciting to explore anomalous diffusion, ballistic
motion, and billiard dynamics. Oscillatory motion in a confining trap adds a
modulated component to the velocity correlation function, which is also
measurable as periodic revivals of spatial structures.

Various matter-wave phenomena can find their analogue in polariton diffusion,
as diffusion manifests the diffraction equation in imaginary time. Thus, a
speckle field of 'traps' that locally depolarize the dark state relates to the
Anderson problem in one or two dimensions, and is measurable spectrally and
spatially. Here one can extend the study to the sub-diffractive, sub-diffusive
($\nabla^{4}$) evolution \cite{HerreroPRE2006}\textsl{ }by controlling the
slow-light parameters. Identifying the transverse modes of either ordered or
disordered configurations is an important, instructive stage for understanding
these systems \cite{GenackNature2011}. Extensions to the nonlinear realm can
be performed with diffusion and diffraction manipulation in Raman 4-wave
mixing schemes, which will further allow optical conjugation and gain
\cite{LettNature2009,Katzir2012}. These promising avenues, which represent a
subset of what is currently being explored in this exciting field, are not
only of fundamental interest, but could also have a profound impact on future
quantum-technology applications.

\bigskip We thank R. Pugatch for years of inspiring collaboration and
gratefully acknowledge discussions with P. London, Y. Sagi, and D. Yanekelv.
OF acknowledges support from the Harvard Quantum Optics Center. ND
acknowledges support by the ISF, DIP and Minerva foundations.%

%

\bibliographystyle{apsrmp}
\bibliography{references}

\end{document}